%% file: belle2.tex
 \documentclass[preprint,prd,tightenlines,superscriptaddress]{revtex4-1}

\usepackage{graphicx} 
\usepackage{dcolumn}  
\usepackage{colordvi}
\usepackage{color}
\usepackage{epstopdf}
\usepackage{pstricks}
\usepackage{amssymb}
\usepackage{url}
\graphicspath{{ps}}
\usepackage{hyperref}
\usepackage{tabularx}
\usepackage{multirow}
\usepackage{units}
\usepackage{upgreek}
\usepackage{siunitx}
\usepackage{hyphenat}
\usepackage{subfigure}
\usepackage[italic]{hepnames}

\renewcommand{\PBzero}{\ensuremath{\HepParticle{\PB}{}{}^0}\xspace}

\begin{document}

\def\belletwo {\it {Belle II}}

\clubpenalty = 10000  
\widowpenalty = 10000 

\vspace*{-3\baselineskip}
\resizebox{!}{3cm}{\includegraphics{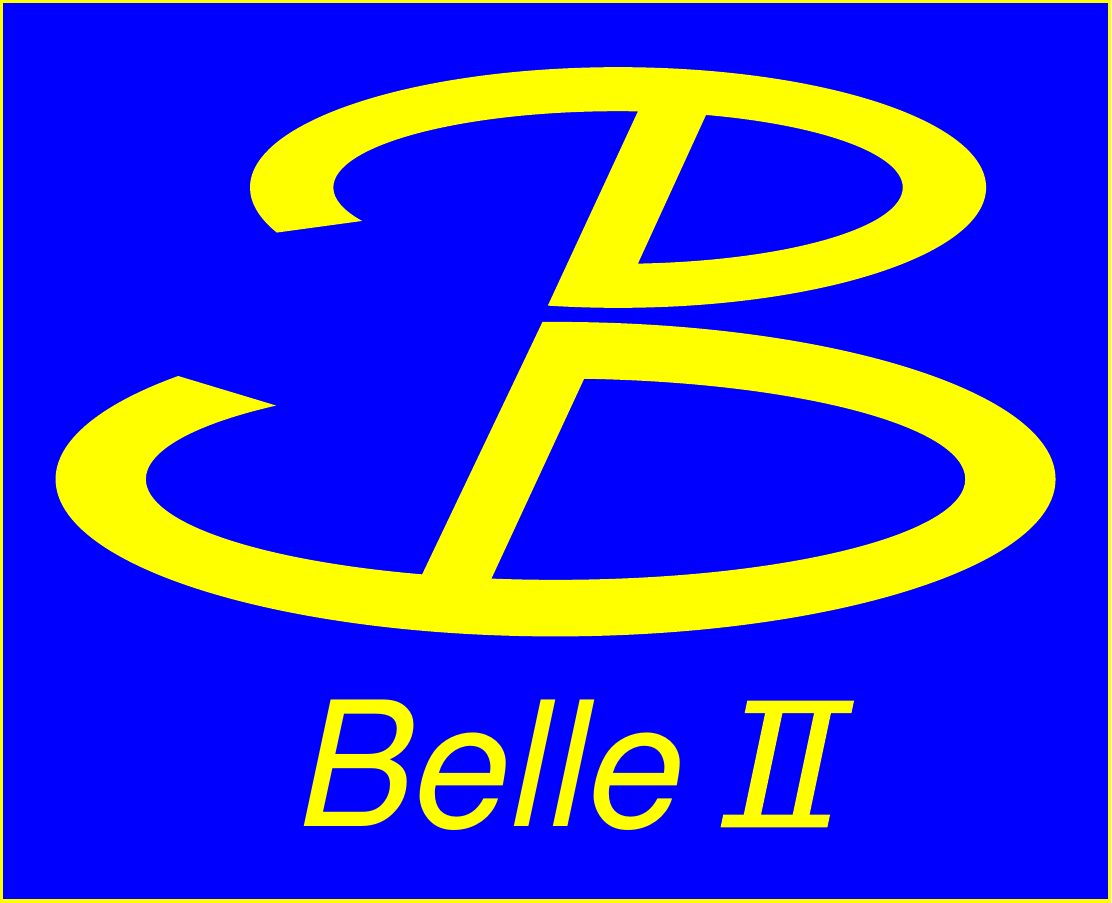}}

\vspace*{-5\baselineskip}
\begin{flushright}
BELLE2-CONF-PH-2021-006\\
\today
\end{flushright}
\quad\\[0.5cm]
\title {Measurements of branching fractions and direct {\it CP}-violating asymmetries in $B^+ \to K^+ \pi^0~\mbox{and}~\pi^+ \pi^0$ decays using 2019 and 2020 Belle II data}

\input{contents/Author_Extra}
\begin{abstract}
We report measurements of branching fractions ($\mathcal B$) and direct {\it CP}-violating asymmetries ($\mathcal A_{\it CP}$) for the decays $B^+\to K^+\pi^0$ and $B^+ \to \pi^+\pi^0$ reconstructed with the Belle II detector in a sample of asymmetric-energy electron-positron collisions at the $\Upsilon(4S)$ resonance corresponding to 62.8 $\text{fb}^{-1}$ of integrated luminosity.
The results are $\mathcal{B}(B^+ \to K^+\pi^0) = [11.9 ^{+1.1}_{-1.0} (\rm stat)\pm 1.6(\rm syst)]\times 10^{-6}$, $\mathcal{B}(B^+ \to \pi^+\pi^0) = [5.5 ^{+1.0}_{-0.9} (\rm stat)\pm 0.7(\rm syst)]\times 10^{-6}$, $\mathcal A_{\it CP}(B^+ \to K^+\pi^0) = -0.09 \pm 0.09 (\rm stat)\pm 0.03(\rm syst)$, and $\mathcal A_{\it CP}(B^+ \to \pi^+\pi^0) = -0.04 \pm 0.17 (\rm stat)\pm 0.06(\rm syst)$. The results are consistent with previous measurements and show a detector performance comparable with early Belle performance.

\end{abstract}


\maketitle

{\renewcommand{\thefootnote}{\fnsymbol{footnote}}}
\setcounter{footnote}{0}
\clearpage

\input{contents/Introduction_Motivation}
\input{contents/Detector}
\input{contents/Sample}
\input{contents/Particle_Selection_and_Reconstruction}
\input{contents/Background}
\input{contents/Control_Sample}
\input{contents/Determination_of_Signal_Yields}
\input{contents/Measurements_for_Branching_Factions_and_CP_Asymmetry}
\input{contents/Systematic_Error_Extra}
\input{contents/Summary_Conclusion}
\input{contents/acknowledgements}
\appendix
\input{contents/Non_Signal_Enhance}
\clearpage
\bibliography{belle2}
\bibliographystyle{belle2-note}

\end{document}

%% file: contents/Author_Extra.tex
\newcommand{\instCPPM}{Aix Marseille Universit\'{e}, CNRS/IN2P3, CPPM, 13288 Marseille, France}
\newcommand{\instBeihang}{Beihang University, Beijing 100191, China}
\newcommand{\instBNL}{Brookhaven National Laboratory, Upton, New York 11973, U.S.A.}
\newcommand{\instBINP}{Budker Institute of Nuclear Physics SB RAS, Novosibirsk 630090, Russian Federation}
\newcommand{\instCMU}{Carnegie Mellon University, Pittsburgh, Pennsylvania 15213, U.S.A.}
\newcommand{\instCinvestavIPN}{Centro de Investigacion y de Estudios Avanzados del Instituto Politecnico Nacional, Mexico City 07360, Mexico}
\newcommand{\instPrague}{Faculty of Mathematics and Physics, Charles University, 121 16 Prague, Czech Republic}
\newcommand{\instChiangMai}{Chiang Mai University, Chiang Mai 50202, Thailand}
\newcommand{\instChiba}{Chiba University, Chiba 263-8522, Japan}
\newcommand{\instChonnam}{Chonnam National University, Gwangju 61186, South Korea}
\newcommand{\instConacyt}{Consejo Nacional de Ciencia y Tecnolog\'{\i}a, Mexico City 03940, Mexico}
\newcommand{\instDESY}{Deutsches Elektronen--Synchrotron, 22607 Hamburg, Germany}
\newcommand{\instDuke}{Duke University, Durham, North Carolina 27708, U.S.A.}
\newcommand{\instITAR}{Institute of Theoretical and Applied Research (ITAR), Duy Tan University, Hanoi 100000, Vietnam}
\newcommand{\instRomaENEA}{ENEA Casaccia, I-00123 Roma, Italy}
\newcommand{\instEri}{Earthquake Research Institute, University of Tokyo, Tokyo 113-0032, Japan}
\newcommand{\instJuelich}{Forschungszentrum J\"{u}lich, 52425 J\"{u}lich, Germany}
\newcommand{\instFuJen}{Department of Physics, Fu Jen Catholic University, Taipei 24205, Taiwan}
\newcommand{\instFudan}{Key Laboratory of Nuclear Physics and Ion-beam Application (MOE) and Institute of Modern Physics, Fudan University, Shanghai 200443, China}
\newcommand{\instGoettingen}{II. Physikalisches Institut, Georg-August-Universit\"{a}t G\"{o}ttingen, 37073 G\"{o}ttingen, Germany}
\newcommand{\instGifu}{Gifu University, Gifu 501-1193, Japan}
\newcommand{\instSOKENDAI}{The Graduate University for Advanced Studies (SOKENDAI), Hayama 240-0193, Japan}
\newcommand{\instGyeongsang}{Gyeongsang National University, Jinju 52828, South Korea}
\newcommand{\instHanyang}{Department of Physics and Institute of Natural Sciences, Hanyang University, Seoul 04763, South Korea}
\newcommand{\instKEK}{High Energy Accelerator Research Organization (KEK), Tsukuba 305-0801, Japan}
\newcommand{\instJPARC}{J-PARC Branch, KEK Theory Center, High Energy Accelerator Research Organization (KEK), Tsukuba 305-0801, Japan}
\newcommand{\instHSE}{National Research University Higher School of Economics, Moscow 101000, Russian Federation}
\newcommand{\instIISER}{Indian Institute of Science Education and Research Mohali, SAS Nagar, 140306, India}
\newcommand{\instIITBhubaneswar}{Indian Institute of Technology Bhubaneswar, Satya Nagar 751007, India}
\newcommand{\instIITGuwahati}{Indian Institute of Technology Guwahati, Assam 781039, India}
\newcommand{\instIITHyderabad}{Indian Institute of Technology Hyderabad, Telangana 502285, India}
\newcommand{\instIITMadras}{Indian Institute of Technology Madras, Chennai 600036, India}
\newcommand{\instIndiana}{Indiana University, Bloomington, Indiana 47408, U.S.A.}
\newcommand{\instIHEPRussia}{Institute for High Energy Physics, Protvino 142281, Russian Federation}
\newcommand{\instHEPHYVienna}{Institute of High Energy Physics, Vienna 1050, Austria}
\newcommand{\instHiroshima}{Hiroshima University, Higashi-Hiroshima, Hiroshima 739-8530, Japan}
\newcommand{\instIHEPChina}{Institute of High Energy Physics, Chinese Academy of Sciences, Beijing 100049, China}
\newcommand{\instIPP}{Institute of Particle Physics (Canada), Victoria, British Columbia V8W 2Y2, Canada}
\newcommand{\instIOP}{Institute of Physics, Vietnam Academy of Science and Technology (VAST), Hanoi, Vietnam}
\newcommand{\instIFIC}{Instituto de Fisica Corpuscular, Paterna 46980, Spain}
\newcommand{\instFrascati}{INFN Laboratori Nazionali di Frascati, I-00044 Frascati, Italy}
\newcommand{\instNapoliINFN}{INFN Sezione di Napoli, I-80126 Napoli, Italy}
\newcommand{\instPadovaINFN}{INFN Sezione di Padova, I-35131 Padova, Italy}
\newcommand{\instPerugiaINFN}{INFN Sezione di Perugia, I-06123 Perugia, Italy}
\newcommand{\instPisaINFN}{INFN Sezione di Pisa, I-56127 Pisa, Italy}
\newcommand{\instRomaINFN}{INFN Sezione di Roma, I-00185 Roma, Italy}
\newcommand{\instRomaTreINFN}{INFN Sezione di Roma Tre, I-00146 Roma, Italy}
\newcommand{\instTorinoINFN}{INFN Sezione di Torino, I-10125 Torino, Italy}
\newcommand{\instTriesteINFN}{INFN Sezione di Trieste, I-34127 Trieste, Italy}
\newcommand{\instJAEA}{Advanced Science Research Center, Japan Atomic Energy Agency, Naka 319-1195, Japan}
\newcommand{\instMainz}{Johannes Gutenberg-Universit\"{a}t Mainz, Institut f\"{u}r Kernphysik, D-55099 Mainz, Germany}
\newcommand{\instGiessen}{Justus-Liebig-Universit\"{a}t Gie\ss{}en, 35392 Gie\ss{}en, Germany}
\newcommand{\instKarlsruhe}{Institut f\"{u}r Experimentelle Teilchenphysik, Karlsruher Institut f\"{u}r Technologie, 76131 Karlsruhe, Germany}
\newcommand{\instISU}{Iowa State University, Ames, Iowa 50011, U.S.A.}
\newcommand{\instKitasato}{Kitasato University, Sagamihara 252-0373, Japan}
\newcommand{\instKISTI}{Korea Institute of Science and Technology Information, Daejeon 34141, South Korea}
\newcommand{\instKoreaUnivKU}{Korea University, Seoul 02841, South Korea}
\newcommand{\instKSU}{Kyoto Sangyo University, Kyoto 603-8555, Japan}
\newcommand{\instKyungpook}{Kyungpook National University, Daegu 41566, South Korea}
\newcommand{\instLPI}{P.N. Lebedev Physical Institute of the Russian Academy of Sciences, Moscow 119991, Russian Federation}
\newcommand{\instLNNU}{Liaoning Normal University, Dalian 116029, China}
\newcommand{\instLMU}{Ludwig Maximilians University, 80539 Munich, Germany}
\newcommand{\instLuther}{Luther College, Decorah, Iowa 52101, U.S.A.}
\newcommand{\instMNITJaipur}{Malaviya National Institute of Technology Jaipur, Jaipur 302017, India}
\newcommand{\instMPP}{Max-Planck-Institut f\"{u}r Physik, 80805 M\"{u}nchen, Germany}
\newcommand{\instMPGHLL}{Semiconductor Laboratory of the Max Planck Society, 81739 M\"{u}nchen, Germany}
\newcommand{\instMcGill}{McGill University, Montr\'{e}al, Qu\'{e}bec, H3A 2T8, Canada}
\newcommand{\instMEPhI}{Moscow Physical Engineering Institute, Moscow 115409, Russian Federation}
\newcommand{\instNagoya}{Graduate School of Science, Nagoya University, Nagoya 464-8602, Japan}
\newcommand{\instNagoyaIAR}{Institute for Advanced Research, Nagoya University, Nagoya 464-8602, Japan}
\newcommand{\instNagoyaKMI}{Kobayashi-Maskawa Institute, Nagoya University, Nagoya 464-8602, Japan}
\newcommand{\instNaraWu}{Nara Women's University, Nara 630-8506, Japan}
\newcommand{\instNTUTaiwan}{Department of Physics, National Taiwan University, Taipei 10617, Taiwan}
\newcommand{\instNUUTaiwan}{National United University, Miao Li 36003, Taiwan}
\newcommand{\instKrakow}{H. Niewodniczanski Institute of Nuclear Physics, Krakow 31-342, Poland}
\newcommand{\instNiigata}{Niigata University, Niigata 950-2181, Japan}
\newcommand{\instNSU}{Novosibirsk State University, Novosibirsk 630090, Russian Federation}
\newcommand{\instOkinawa}{Okinawa Institute of Science and Technology, Okinawa 904-0495, Japan}
\newcommand{\instOsakaCity}{Osaka City University, Osaka 558-8585, Japan}
\newcommand{\instRCNP}{Research Center for Nuclear Physics, Osaka University, Osaka 567-0047, Japan}
\newcommand{\instPNNL}{Pacific Northwest National Laboratory, Richland, Washington 99352, U.S.A.}
\newcommand{\instPanjab}{Panjab University, Chandigarh 160014, India}
\newcommand{\instPanjabPAU}{Punjab Agricultural University, Ludhiana 141004, India}
\newcommand{\instRIKENMSL}{Meson Science Laboratory, Cluster for Pioneering Research, RIKEN, Saitama 351-0198, Japan}
\newcommand{\instSeoul}{Seoul National University, Seoul 08826, South Korea}
\newcommand{\instSPU}{Showa Pharmaceutical University, Tokyo 194-8543, Japan}
\newcommand{\instSoochow}{Soochow University, Suzhou 215006, China}
\newcommand{\instSoongsil}{Soongsil University, Seoul 06978, South Korea}
\newcommand{\instLjubljanaJSI}{J. Stefan Institute, 1000 Ljubljana, Slovenia}
\newcommand{\instKyiv}{Taras Shevchenko National Univ. of Kiev, Kiev, Ukraine}
\newcommand{\instTata}{Tata Institute of Fundamental Research, Mumbai 400005, India}
\newcommand{\instTUM}{Department of Physics, Technische Universit\"{a}t M\"{u}nchen, 85748 Garching, Germany}
\newcommand{\instTelAviv}{Tel Aviv University, School of Physics and Astronomy, Tel Aviv, 69978, Israel}
\newcommand{\instToho}{Toho University, Funabashi 274-8510, Japan}
\newcommand{\instTohoku}{Department of Physics, Tohoku University, Sendai 980-8578, Japan}
\newcommand{\instTitech}{Tokyo Institute of Technology, Tokyo 152-8550, Japan}
\newcommand{\instTokyoMetropolitan}{Tokyo Metropolitan University, Tokyo 192-0397, Japan}
\newcommand{\instUAS}{Universidad Autonoma de Sinaloa, Sinaloa 80000, Mexico}
\newcommand{\instNapoliUNIV}{Dipartimento di Scienze Fisiche, Universit\`{a} di Napoli Federico II, I-80126 Napoli, Italy}
\newcommand{\instPadovaUNIV}{Dipartimento di Fisica e Astronomia, Universit\`{a} di Padova, I-35131 Padova, Italy}
\newcommand{\instPerugiaUNIV}{Dipartimento di Fisica, Universit\`{a} di Perugia, I-06123 Perugia, Italy}
\newcommand{\instPisaUNIV}{Dipartimento di Fisica, Universit\`{a} di Pisa, I-56127 Pisa, Italy}
\newcommand{\instRomaTreUNIV}{Dipartimento di Matematica e Fisica, Universit\`{a} di Roma Tre, I-00146 Roma, Italy}
\newcommand{\instTorinoUNIV}{Dipartimento di Fisica, Universit\`{a} di Torino, I-10125 Torino, Italy}
\newcommand{\instTriesteUNIV}{Dipartimento di Fisica, Universit\`{a} di Trieste, I-34127 Trieste, Italy}
\newcommand{\instMontreal}{Universit\'{e} de Montr\'{e}al, Physique des Particules, Montr\'{e}al, Qu\'{e}bec, H3C 3J7, Canada}
\newcommand{\instIJCLab}{Universit\'{e} Paris-Saclay, CNRS/IN2P3, IJCLab, 91405 Orsay, France}
\newcommand{\instIPHC}{Universit\'{e} de Strasbourg, CNRS, IPHC, UMR 7178, 67037 Strasbourg, France}
\newcommand{\instAdelaide}{Department of Physics, University of Adelaide, Adelaide, South Australia 5005, Australia}
\newcommand{\instBonn}{University of Bonn, 53115 Bonn, Germany}
\newcommand{\instUBC}{University of British Columbia, Vancouver, British Columbia, V6T 1Z1, Canada}
\newcommand{\instCincinnati}{University of Cincinnati, Cincinnati, Ohio 45221, U.S.A.}
\newcommand{\instFlorida}{University of Florida, Gainesville, Florida 32611, U.S.A.}
\newcommand{\instHawaii}{University of Hawaii, Honolulu, Hawaii 96822, U.S.A.}
\newcommand{\instHeidelberg}{University of Heidelberg, 68131 Mannheim, Germany}
\newcommand{\instLjubljanaUniLJ}{Faculty of Mathematics and Physics, University of Ljubljana, 1000 Ljubljana, Slovenia}
\newcommand{\instLouisville}{University of Louisville, Louisville, Kentucky 40292, U.S.A.}
\newcommand{\instMalaya}{National Centre for Particle Physics, University Malaya, 50603 Kuala Lumpur, Malaysia}
\newcommand{\instLjubljanaUM}{Faculty of Chemistry and Chemical Engineering, University of Maribor, 2000 Maribor, Slovenia}
\newcommand{\instMelbourne}{School of Physics, University of Melbourne, Victoria 3010, Australia}
\newcommand{\instMississippi}{University of Mississippi, University, Mississippi 38677, U.S.A.}
\newcommand{\instUOM}{University of Miyazaki, Miyazaki 889-2192, Japan}
\newcommand{\instPittsburgh}{University of Pittsburgh, Pittsburgh, Pennsylvania 15260, U.S.A.}
\newcommand{\instUSTC}{University of Science and Technology of China, Hefei 230026, China}
\newcommand{\instSAlabama}{University of South Alabama, Mobile, Alabama 36688, U.S.A.}
\newcommand{\instSCarolina}{University of South Carolina, Columbia, South Carolina 29208, U.S.A.}
\newcommand{\instSydney}{School of Physics, University of Sydney, New South Wales 2006, Australia}
\newcommand{\instUTokyo}{Department of Physics, University of Tokyo, Tokyo 113-0033, Japan}
\newcommand{\instIPMU}{Kavli Institute for the Physics and Mathematics of the Universe (WPI), University of Tokyo, Kashiwa 277-8583, Japan}
\newcommand{\instVictoria}{University of Victoria, Victoria, British Columbia, V8W 3P6, Canada}
\newcommand{\instVPI}{Virginia Polytechnic Institute and State University, Blacksburg, Virginia 24061, U.S.A.}
\newcommand{\instWayneState}{Wayne State University, Detroit, Michigan 48202, U.S.A.}
\newcommand{\instYamagata}{Yamagata University, Yamagata 990-8560, Japan}
\newcommand{\instYerevan}{Alikhanyan National Science Laboratory, Yerevan 0036, Armenia}
\newcommand{\instYonsei}{Yonsei University, Seoul 03722, South Korea}
\newcommand{\instZZU}{Zhengzhou University, Zhengzhou 450001, China}
\affiliation{\instCPPM}
\affiliation{\instBeihang}
\affiliation{\instBNL}
\affiliation{\instBINP}
\affiliation{\instCMU}
\affiliation{\instCinvestavIPN}
\affiliation{\instPrague}
\affiliation{\instChiangMai}
\affiliation{\instChiba}
\affiliation{\instChonnam}
\affiliation{\instConacyt}
\affiliation{\instDESY}
\affiliation{\instDuke}
\affiliation{\instITAR}
\affiliation{\instRomaENEA}
\affiliation{\instEri}
\affiliation{\instJuelich}
\affiliation{\instFuJen}
\affiliation{\instFudan}
\affiliation{\instGoettingen}
\affiliation{\instGifu}
\affiliation{\instSOKENDAI}
\affiliation{\instGyeongsang}
\affiliation{\instHanyang}
\affiliation{\instKEK}
\affiliation{\instJPARC}
\affiliation{\instHSE}
\affiliation{\instIISER}
\affiliation{\instIITBhubaneswar}
\affiliation{\instIITGuwahati}
\affiliation{\instIITHyderabad}
\affiliation{\instIITMadras}
\affiliation{\instIndiana}
\affiliation{\instIHEPRussia}
\affiliation{\instHEPHYVienna}
\affiliation{\instHiroshima}
\affiliation{\instIHEPChina}
\affiliation{\instIPP}
\affiliation{\instIOP}
\affiliation{\instIFIC}
\affiliation{\instFrascati}
\affiliation{\instNapoliINFN}
\affiliation{\instPadovaINFN}
\affiliation{\instPerugiaINFN}
\affiliation{\instPisaINFN}
\affiliation{\instRomaINFN}
\affiliation{\instRomaTreINFN}
\affiliation{\instTorinoINFN}
\affiliation{\instTriesteINFN}
\affiliation{\instJAEA}
\affiliation{\instMainz}
\affiliation{\instGiessen}
\affiliation{\instKarlsruhe}
\affiliation{\instISU}
\affiliation{\instKitasato}
\affiliation{\instKISTI}
\affiliation{\instKoreaUnivKU}
\affiliation{\instKSU}
\affiliation{\instKyungpook}
\affiliation{\instLPI}
\affiliation{\instLNNU}
\affiliation{\instLMU}
\affiliation{\instLuther}
\affiliation{\instMNITJaipur}
\affiliation{\instMPP}
\affiliation{\instMPGHLL}
\affiliation{\instMcGill}
\affiliation{\instMEPhI}
\affiliation{\instNagoya}
\affiliation{\instNagoyaIAR}
\affiliation{\instNagoyaKMI}
\affiliation{\instNaraWu}
\affiliation{\instNTUTaiwan}
\affiliation{\instNUUTaiwan}
\affiliation{\instKrakow}
\affiliation{\instNiigata}
\affiliation{\instNSU}
\affiliation{\instOkinawa}
\affiliation{\instOsakaCity}
\affiliation{\instRCNP}
\affiliation{\instPNNL}
\affiliation{\instPanjab}
\affiliation{\instPanjabPAU}
\affiliation{\instRIKENMSL}
\affiliation{\instSeoul}
\affiliation{\instSPU}
\affiliation{\instSoochow}
\affiliation{\instSoongsil}
\affiliation{\instLjubljanaJSI}
\affiliation{\instKyiv}
\affiliation{\instTata}
\affiliation{\instTUM}
\affiliation{\instTelAviv}
\affiliation{\instToho}
\affiliation{\instTohoku}
\affiliation{\instTitech}
\affiliation{\instTokyoMetropolitan}
\affiliation{\instUAS}
\affiliation{\instNapoliUNIV}
\affiliation{\instPadovaUNIV}
\affiliation{\instPerugiaUNIV}
\affiliation{\instPisaUNIV}
\affiliation{\instRomaTreUNIV}
\affiliation{\instTorinoUNIV}
\affiliation{\instTriesteUNIV}
\affiliation{\instMontreal}
\affiliation{\instIJCLab}
\affiliation{\instIPHC}
\affiliation{\instAdelaide}
\affiliation{\instBonn}
\affiliation{\instUBC}
\affiliation{\instCincinnati}
\affiliation{\instFlorida}
\affiliation{\instHawaii}
\affiliation{\instHeidelberg}
\affiliation{\instLjubljanaUniLJ}
\affiliation{\instLouisville}
\affiliation{\instMalaya}
\affiliation{\instLjubljanaUM}
\affiliation{\instMelbourne}
\affiliation{\instMississippi}
\affiliation{\instUOM}
\affiliation{\instPittsburgh}
\affiliation{\instUSTC}
\affiliation{\instSAlabama}
\affiliation{\instSCarolina}
\affiliation{\instSydney}
\affiliation{\instUTokyo}
\affiliation{\instIPMU}
\affiliation{\instVictoria}
\affiliation{\instVPI}
\affiliation{\instWayneState}
\affiliation{\instYamagata}
\affiliation{\instYerevan}
\affiliation{\instYonsei}
\affiliation{\instZZU}
  \author{F.~Abudin{\'e}n}\affiliation{\instTriesteINFN} 
  \author{I.~Adachi}\affiliation{\instKEK}\affiliation{\instSOKENDAI} 
  \author{R.~Adak}\affiliation{\instFudan} 
  \author{K.~Adamczyk}\affiliation{\instKrakow} 
  \author{P.~Ahlburg}\affiliation{\instBonn} 
  \author{J.~K.~Ahn}\affiliation{\instKoreaUnivKU} 
  \author{H.~Aihara}\affiliation{\instUTokyo} 
  \author{N.~Akopov}\affiliation{\instYerevan} 
  \author{A.~Aloisio}\affiliation{\instNapoliUNIV}\affiliation{\instNapoliINFN} 
  \author{F.~Ameli}\affiliation{\instRomaINFN} 
% ...

%% file: contents/Introduction_Motivation.tex
\section{Introduction and motivation}

\begin{figure}[htb]
\centering
\includegraphics[width=0.475\textwidth]{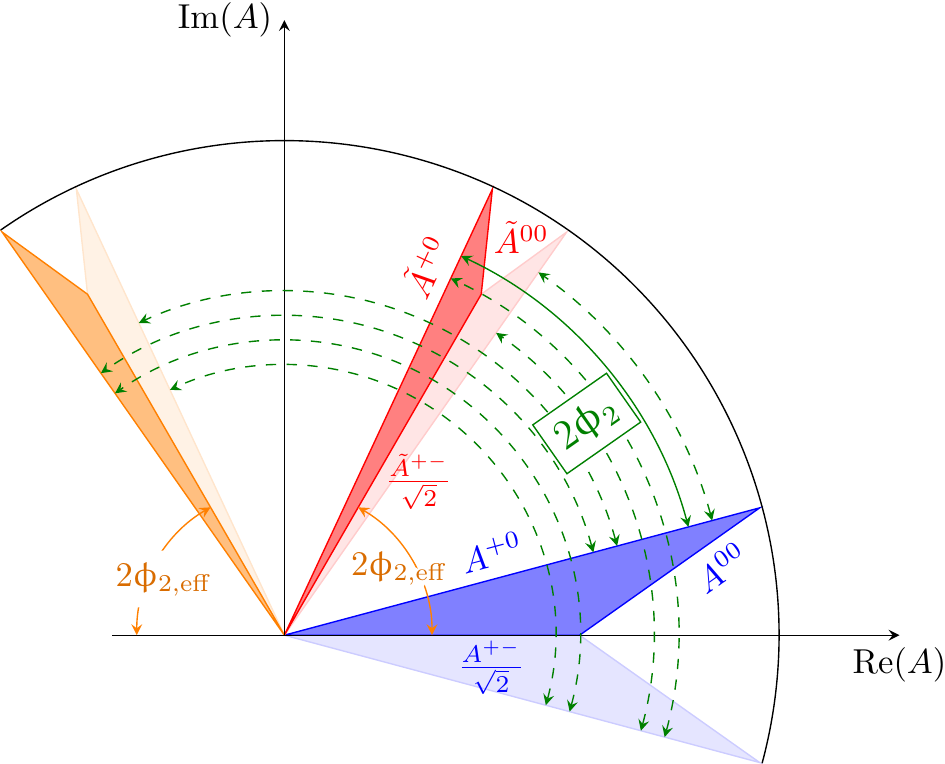}
\caption{Geometrical representation of the isospin triangular relations in the complex plane of $B^{i+j}\to h^i h^j$ amplitudes ($A^{ij}$). The blue and the red shaded areas correspond to the isospin triangles. The angle between the {\it CP} conjugate charged amplitudes $A^{+-}$ and $\tilde{A}^{+-}$ corresponds to twice the weak phase $\alpha_{\text{eff}}/\phi_{2,\text{eff}}$ (orange solid curves). The angle between the {\it CP} conjugate charged amplitudes $A^{+0}$ and $\tilde{A}^{+0}$ corresponds to twice the CKM angle $\alpha/\phi_2$ (green solid curve). The other triangles with lighter shades represent the mirror solutions allowed by the discrete ambiguities associated with the isospin relationships, with the corresponding values for $\alpha/\phi_2$ represented by the green dashed curves.}
\label{fig:a}
\end{figure}

The study of charmless $B$ decays is a keystone of the worldwide flavor program. Processes mediated by $b\to sq\overline{q}$ 
transitions probe contributions of non-standard-model dynamics in loop decay-amplitudes.  However, reliable extraction of weak phases and unambiguous interpretation of measurements involving these amplitudes is spoiled by large hadronic uncertainties, which are rarely tractable in perturbative calculations. Appropriately chosen combinations of measurements from decay modes related by flavor symmetries are used to significantly reduce the impact of such unknowns. An especially  fruitful approach consists in combining measurements from decays related by isospin symmetries. This approach has been proposed to address the so-called $K\pi$ {\it puzzle}, a long-standing anomaly associated with the significant difference between direct {\it CP}-violating asymmetries observed in $B^0 \to K^+\pi^-$ and $B^+ \to K^+\pi^0$ decays~\cite{Gronau:1999}. The asymmetries are expected to be equal at leading order in electroweak perturbation theory, as the two decays differ only by the \emph{spectator} quark. The isospin sum-rule 
\begin{equation}
\begin{aligned}
I_{K\pi} = \mathcal{A}_{K^+\pi^-} + \mathcal{A}_{K^0\pi^+}\frac{\mathcal{B}(K^0\pi^+)}{\mathcal{B}(K^+\pi^-)}\frac{\tau_{B^0}}{\tau_{B^+}} - 2\mathcal{A}_{K^+\pi^0}\frac{\mathcal{B}(K^+\pi^0)}{\mathcal{B}(K^+\pi^-)}\frac{\tau_{B^0}}{\tau_{B^+}} - 2\mathcal{A}_{K^0\pi^0}\frac{\mathcal{B}(K^0\pi^0)}{\mathcal{B}(K^+\pi^-)}~,
\label{eq:SumRule}
\end{aligned}
\end{equation}
properly accounts for subleading amplitudes by combining the branching fractions ($\mathcal{B}$) and direct {\it CP} asymmetries ($\mathcal{A}$) of $B^+(B^0)$ decays to all four final states $K^+ \pi^-$, $K^0\pi^+$, $K^+\pi^0$, and $K^0 \pi^0$, and the lifetime ($\tau$) ratio between $B^+$ and $B^0$. This rule offers a stringent null test of the standard model (SM), which 
predicts $I_{K\pi} = 0$ in the limit of exact isospin symmetry and no electroweak penguin (EWP) contributions, and with an uncertainty much below $1\%$ when including SM EWP amplitudes~\cite{Gronau:2005kz,Gershon:2007}. For $B^+ \to \pi^+\pi^0$ decay, the branching fraction is an ingredient (Figure~\ref{fig:a}) for an isospin-based determination of $\alpha/\phi_2$ based on $B \to \pi\pi$ decays~\cite{Gronau:1990ka}.
Belle~II has the unique capability of studying jointly, and within a consistent experimental environment, all relevant final states.

The Belle~II experiment, complete with its vertex detector, started its collision operations in March 2019 and is currently collecting data.  The sample of electron-positron collisions used in this work corresponds to an integrated luminosity of $62.8\,\si{fb^{-1}}$
and was collected at the $\Upsilon(4{\rm S})$~resonance.

We report updated Belle~II measurements of branching fractions and direct {\it CP}-violating asymmetries in $B^+ \to K^+ \pi^0$  and  $B^+ \to \pi^+ \pi^0$  decays. These results supersede those in Ref~\cite{Charmless:2020} using more data and an improved analysis. 
All analysis procedures are first developed and finalized in simulated data and then tested on the data used for summer 2020 results~\cite{Charmless:2020} prior to the application on the full data set.

%% file: contents/Detector.tex
\section{The Belle II detector}
\label{sec:Detectors}
Belle~II is a $4\pi$ particle-physics spectrometer~\cite{Kou:2018nap, Abe:2010sj}, designed to reconstruct the products of electron-positron collisions produced by the SuperKEKB asymmetric-energy collider~\cite{Akai:2018mbz}, located at the KEK laboratory in Tsukuba, Japan. Belle~II comprises several subdetectors arranged around the interaction space-point in a cylindrical geometry. The innermost subdetector is the vertex detector, which uses position-sensitive silicon layers to sample the trajectories of charged particles (tracks) in the vicinity of the interaction region in order to extrapolate the decay positions of their long-lived parent particles. The vertex detector includes two inner layers of silicon pixel sensors and four outer layers of silicon microstrip sensors. The second pixel layer is currently incomplete and covers only a small portion of azimuthal angle. Charged-particle momenta and charges are measured by a large-radius, helium-ethane, small-cell central drift chamber, which also offers charged-particle-identification information through a measurement of particles' energy-loss by specific ionization. A Cherenkov-light angle and time-of-propagation detector surrounding the chamber provides charged-particle identification in the central detector volume, supplemented by proximity-focusing, aerogel, ring-imaging Cherenkov detectors in the forward regions. A CsI(Tl)-crystal electromagnetic calorimeter allows for energy measurements of electrons and photons.  A solenoid surrounding the calorimeter generates a uniform axial 1.5\,T magnetic field filling its inner volume. Layers of plastic scintillator and resistive-plate chambers, interspersed between the
magnetic flux-return iron plates, allow for identification of $K^0_{\rm L}$ and muons.
The subdetectors most relevant for this work are the silicon vertex detector, the tracking drift chamber, the particle-identification detectors, and the electromagnetic calorimeter.

%% file: contents/Sample.tex
\section{Data and simulation}
\label{sec:sample}
We use all 2019--2020 $\Upsilon$(4S) collected on $\Upsilon(4S)$ resonance until July 1, 2020 corresponding to an integrated luminosity of $62.8\,\si{fb^{-1}}$.
~All events are required to satisfy initial loose data-skim selections, based on total energy and charged-particle multiplicity in the event, targeted at reducing sample sizes to a manageable level with negligible impact on signal efficiency. All data are processed using the Belle~II analysis software~\cite{Kuhr:2018lps}.

We use generic simulated data to optimize the event selection and compare the distributions observed in experimental data with expectations. We use signal-only simulated data to model relevant signal features for fits and determine selection efficiencies. 
Generic simulation consists of Monte Carlo samples that include $B^0\overline{B}^0$, $B^+B^-$, $u\overline{u}$, $d\overline{d}$, $s\overline{s}$, and $c\overline{c}$ processes in realistic proportions and corresponding in size to more than ten times the experimental data.



%% file: contents/Particle_Selection_and_Reconstruction.tex
\section{Event selection and candidate reconstruction}
\label{sec:Selection}
Candidate charged particles are required to be reconstructed in the full polar-angle acceptance ($17^{\circ}<\theta<150^{\circ}$) and originate at the interaction point with longitudinal displacement $|dz| < 3.0$ cm and radial displacement $|dr| < 0.5$ cm to reduce beam-background-induced tracks. Charged kaons and pions are identified using the $dE/dx$ information from the drift chamber and particle identification in the Cherenkov detectors.
~For a 2 GeV/$c$ charged kaon/pion, the identification efficiency is $88\%/91\%$ while $9\%/11\%$ pions/kaons are misidentified as kaons/pions. Candidate neutral pions are reconstructed in their diphoton decays.
Photon pairs with masses between 0.105 and 0.150 $\text{GeV}/c^2$ are selected as $\pi^0$ candidates. 
To reduce the combinatorial background from soft photons, the energy of each candidate photon is required to exceed 22.5 MeV in the forward endcap region and 20 MeV in the barrel and the backward endcap regions. Furthermore, the magnitude of the cosine of the angle between the photon direction in the $\pi^0$ rest frame and the $\pi^0$ direction in the laboratory frame, is required to be less than 0.98. Photon pairs are kinematically fitted to the known $\pi^0$ mass based on their measured energies, directions, and associated covariance matrices. 

Candidate $B$ mesons are selected by pairing a charged kaon or pion candidate with a $\pi^0$ candidate, and are identified using the energy difference, $\Delta E \equiv E^*_B -E^*_{\text{beam}}$ and the modified beam energy-constrained mass, $M_{bc} \equiv \sqrt{{E^{*2}_{\text{beam}}}/c^4 - \vert \Vec{p}^*_{h^{\pm}}/c + \Vec{p}^*(\gamma \gamma)_{\pi^0}/c\vert^2}$ with $\Vec{p}^*(\gamma \gamma)_{\pi^0} = \sqrt{({E^{*2}_{\text{beam}}}/c^2 - {E^{*2}_{h^\pm}}/c^2) - (m^*(\gamma \gamma)_{\pi^0}c)^2} \cdot \hat{p}^*(\gamma \gamma)_{\pi^0}$, where $E^*_{\text{beam}}$ is the run-dependent beam energy, $m^*(\gamma \gamma)_{\pi^0}$ is the reconstructed mass of $\pi^0$ candidates after kinematical fit, and $E^*_{B (h^{\pm})}$, ${p}^*_{h^{\pm}}$, and ${p}^*(\gamma \gamma)_{\pi^0}$ are the reconstructed energy and momentum of $B$, $h^{\pm}$, and $\pi^0$ candidates in the center-of-mass frame, respectively. $B$ candidates with $M_{bc} > 5.2$ $\text{GeV}/c^2$ and $\vert \Delta E \vert < 0.3$ GeV are retained for further analysis. 

%% file: contents/Background.tex
\section{Background reduction}
\begin{figure}[t]
 \centering
    \centering
    \subfigure{\includegraphics[width=0.49\textwidth]{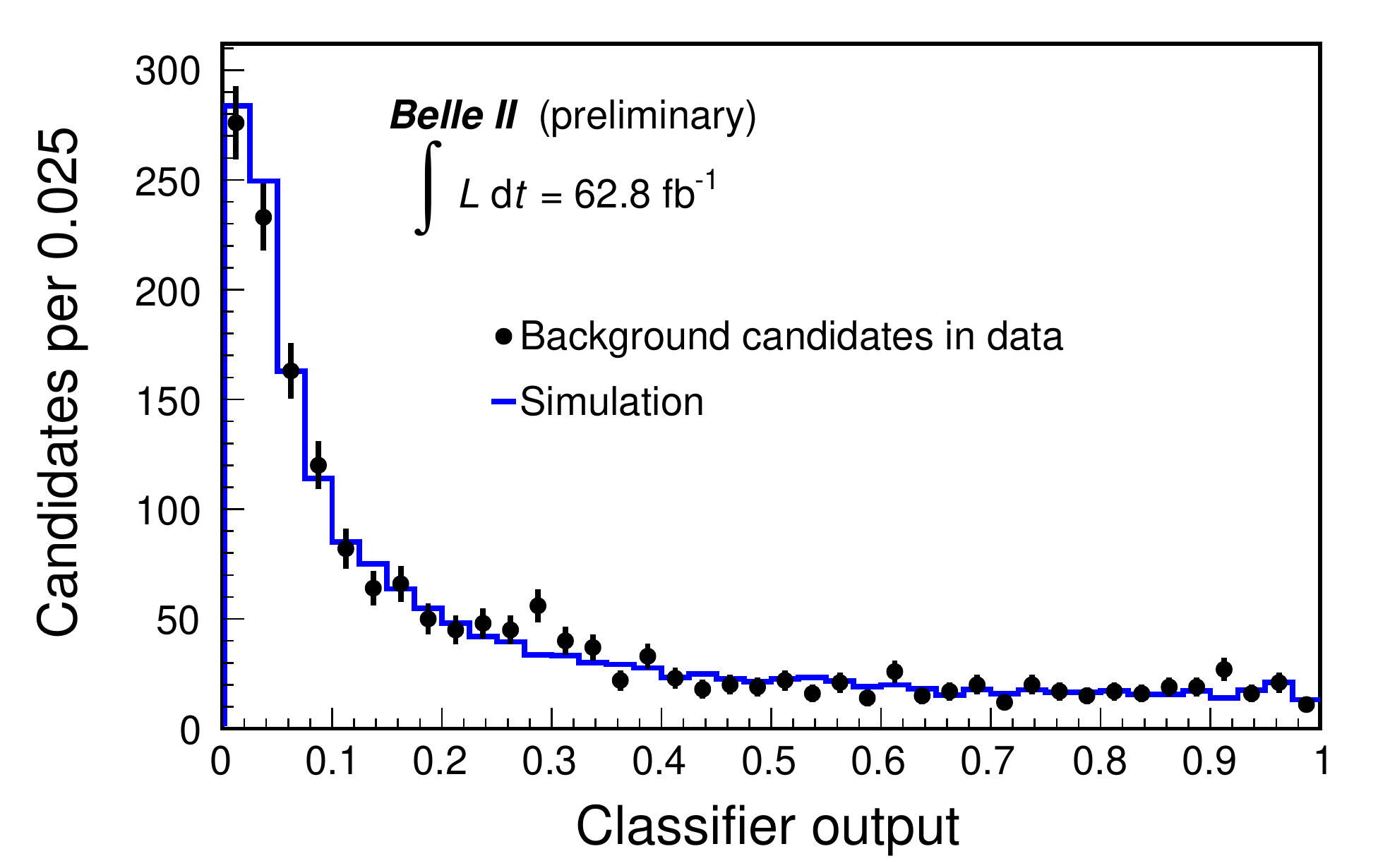}} \hfill
    \subfigure{\includegraphics[width=0.49\textwidth]{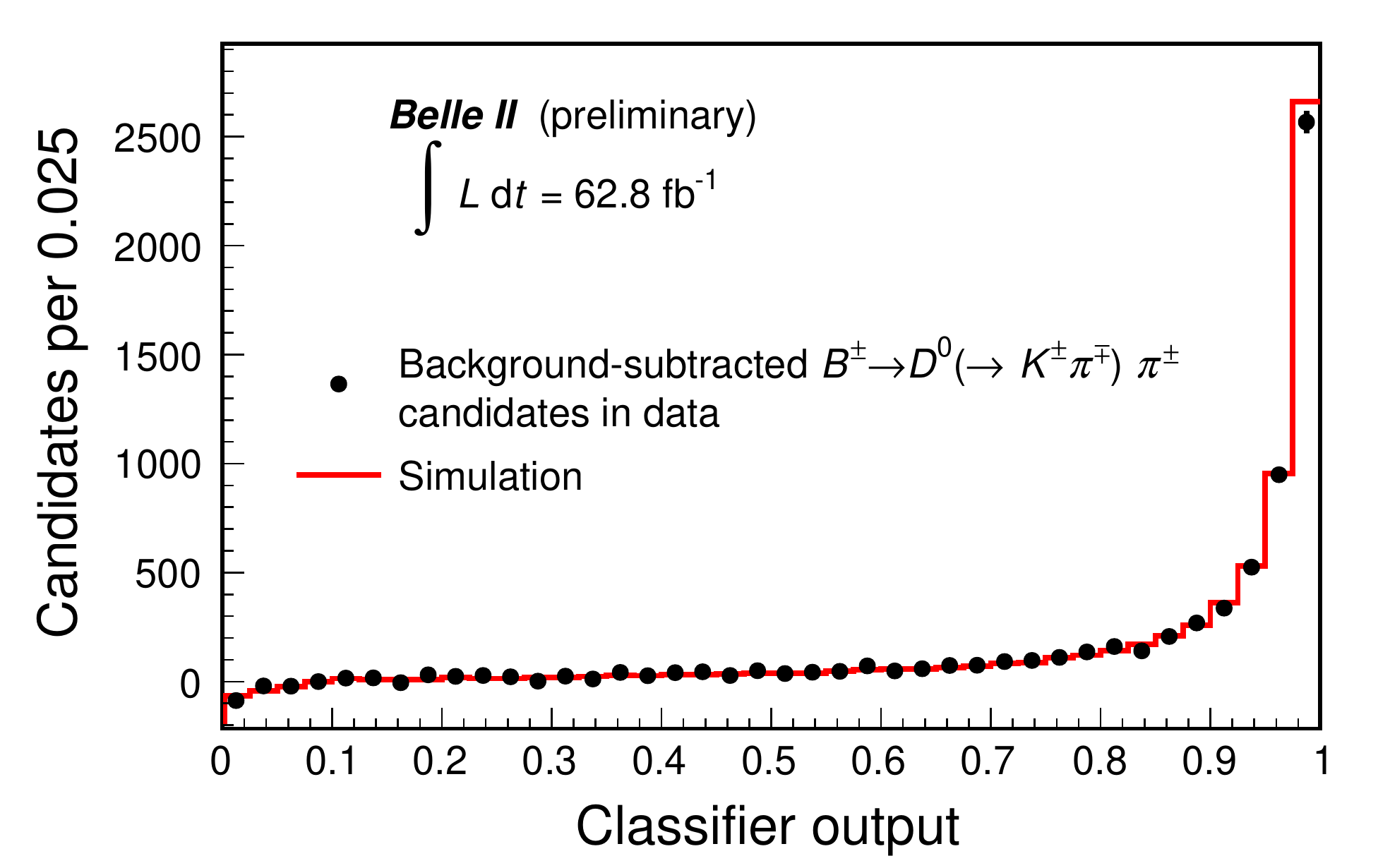}} \\
 \caption{Data-simulation comparison of the output of the boosted decision-tree classifier on (left) sideband and (right) side-band-subtracted $\PBplus\to\overline{D}^{0}(\to \PKp\Pgpm)\,\Pgpp$~candidates in the signal region.}
 \label{fig:outputData_Kpi}
\end{figure}

The main challenge in reconstructing charmless $B$ decay is the large contamination from continuum background. 
We use a binary boosted decision-tree classifier that combines nonlinearly 39 variables known to provide statistical discrimination between $B$-meson signals and continuum background and to be loosely correlated with $M_{bc}$ and $\Delta E$. The variables are quantities associated with event topology (global and signal-only angular configurations), flavor-tagger information, vertex separation and uncertainty information, and kinematic-fit quality information. We train the classifier to identify statistically significant signal and background features using unbiased simulated samples. We validate the input and output distributions of the classifier by comparing control-sample data with simulation. Figure~\ref{fig:outputData_Kpi} shows the distribution of the output for \mbox{$\PBplus\to\overline{D}^{0}(\to \PKp\Pgpm)\,\Pgpp$}~candidates reconstructed in data and simulation. No inconsistency is observed.

To suppress the continuum background effectively, we scan various requirement on the classifier output to search for the criterion that maximises the figure of merit ${\rm S}/\sqrt{{\rm S}+{\rm B}}$, where $S$ and $B$ are the estimated numbers of simulated signal and background candidates in the signal region. The optimal requirement has a signal efficiency and a background retention of $54.6\%$ and $0.7\%$ for the $K^+ \pi^0$ final state, respectively. For the $\pi^+ \pi^0$ final state, the signal efficiency is $44.0\%$, with $0.4\%$ background retention.

More than one signal candidate is reconstructed in about $5\%$ of events. For those events, we keep only the candidate with the best $h^+\pi^0$ kinematic-fit quality. 
A fraction of nonsignal $B$ decays survive the selection, dominated by misidentified $B^+\to \pi^+\pi^0$ and $B^+\to K^+\pi^0$ (feed-across) decays, which are estimated from simulation.
Typical feed-across backgrounds have signal-like shapes in $M_{bc}$ and $\Delta E$, except for a $\Delta E$ shift due to the incorrect mass assignment. 

A significant fraction of background comes from other $B$ decays due to misreconstruction or partial reconstruction of the final state, {\it e.g.}, $B^0 \to K^+\rho^-(\to \pi^- \pi^0)$ or $B^0 \to \pi^+\pi^- \pi^0$. These decays peak like the signal in the $M_{bc}$ distribution and show broader $\Delta E$ peaks shifted toward negative values.

%% file: contents/Control_Sample.tex

\section{Corrections to simulation from control data}
We use the decay $B^+ \to \overline{D}^0(\to K^+\pi^-\pi^0)\pi^+$ to study a bias in the photon-energy calibration, which gives a $\mathcal{O}(10)$\,MeV shift of the signal peak in the $\Delta E$ distribution; we determined corrections for data-simulation discrepancies in the 
signal shape.

Final-state particle selections similar to that of signal decay are applied, with a stringent requirement on the $\pi^0$ momentum ($>1.5$ GeV$/c$). To select kinematic properties as close as possible to that of the signal $\pi^0$.
The correction factors for the simulation are determined by a fit to the $M_{bc}$ and $\Delta E$ distribution of the control sample. 
The signal is modelled with a sum of two Gaussian functions in $M_{bc}$ and a Crystal-Ball function in $\Delta E$. Shape parameters are determined from simulation.
Independent mean shifts and width scale-factors  for $M_{bc}$ and $\Delta E$ are determined from the fit to data as correction factors.
From simulation we estimate a $10\%$ fraction of B decay background coming from particle misidentification, which has shapes similar to the signal in $M_{bc}$ and $\Delta E$. The fraction of this background is Gaussian-constrained to its prediction from simulation and known branching fractions.
Other $B$ decay background from partially reconstructed $B$ decays are modelled by using a two-dimensional kernel-density estimator derived from simulation. 
The continuum background is described by an Argus function in $M_{bc}$ and a second-order polynomial in $\Delta E$. 

\begin{figure}[htb]
\centering
\includegraphics[width=0.475\textwidth]{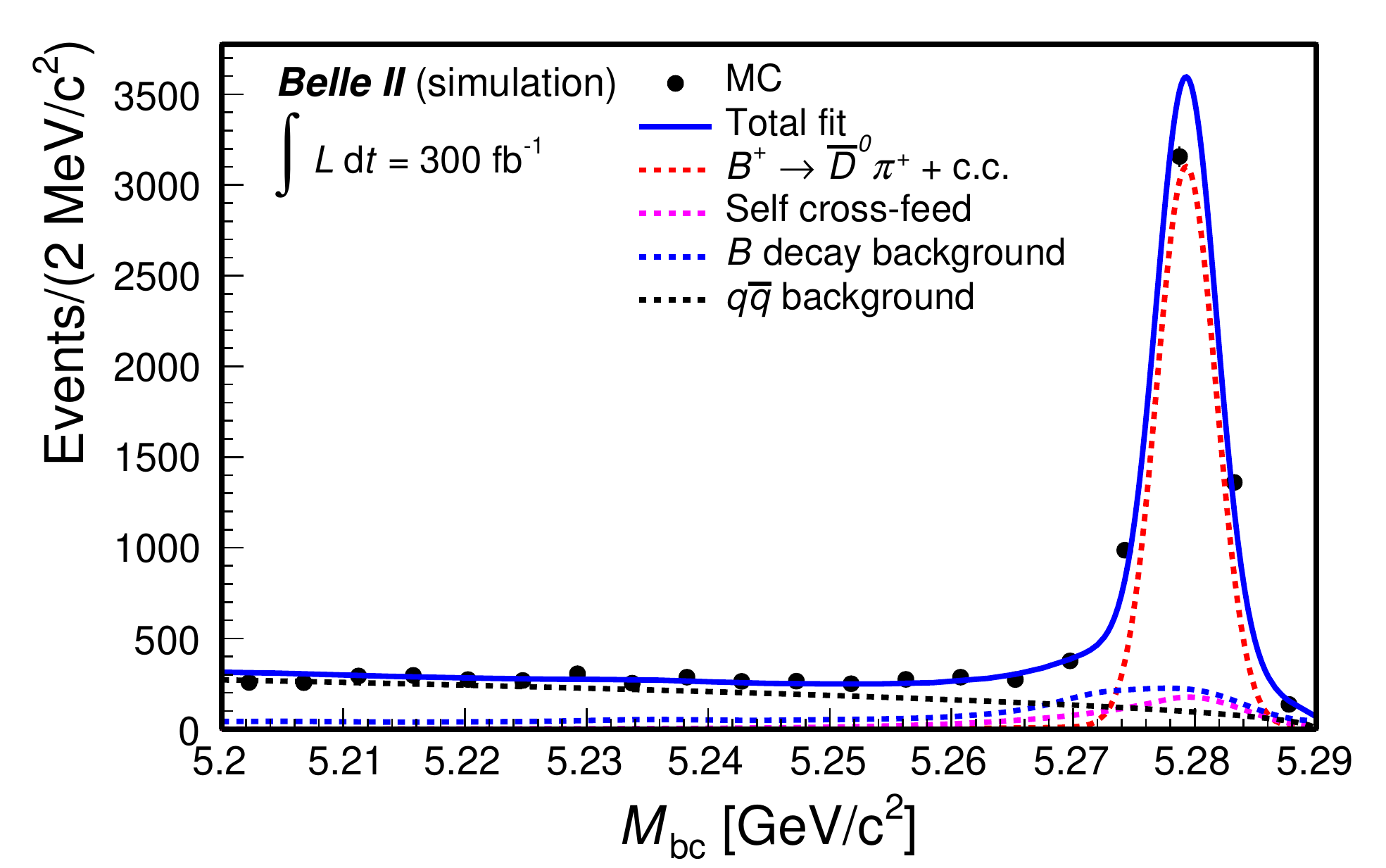}
\includegraphics[width=0.475\textwidth]{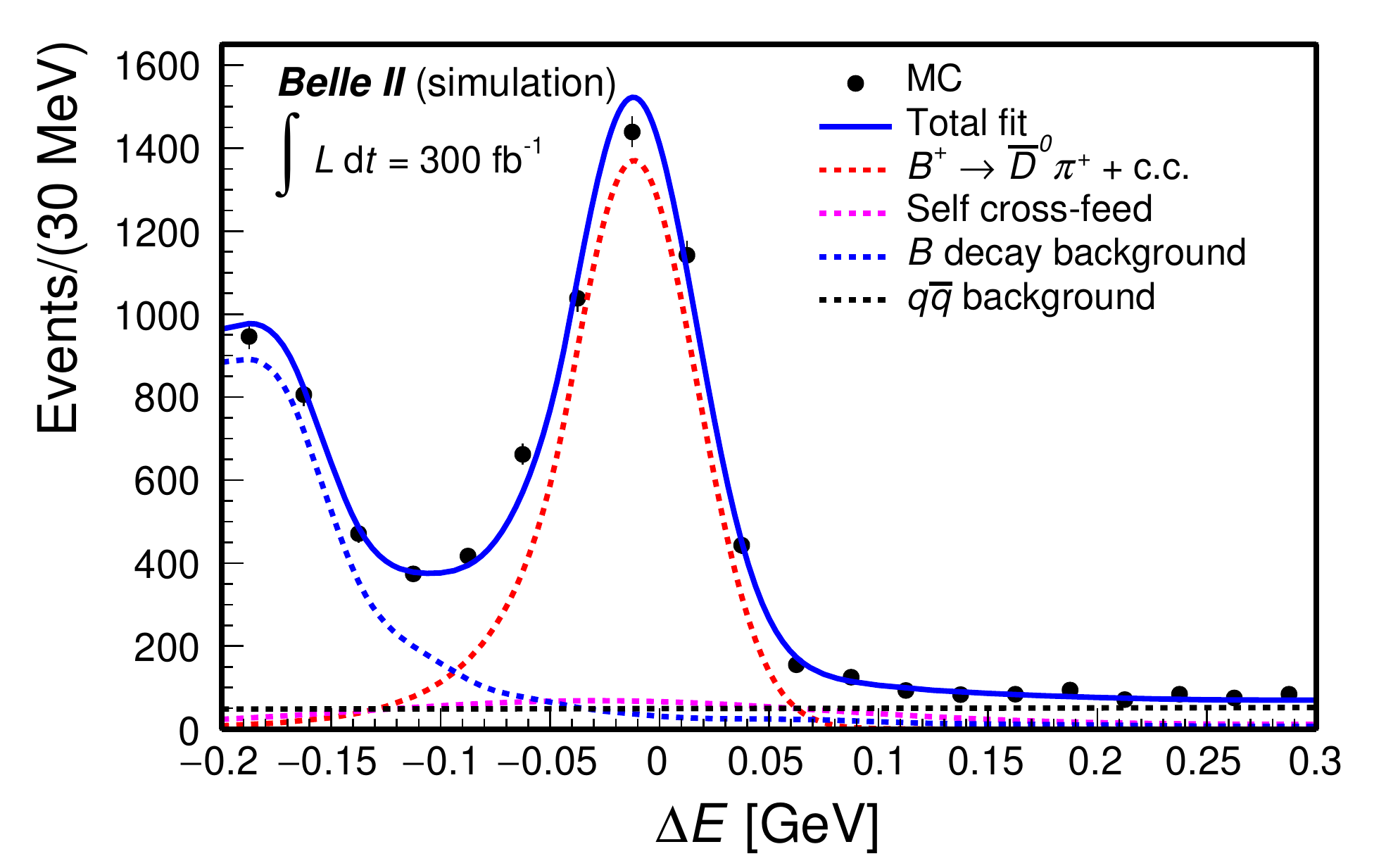}
\includegraphics[width=0.475\textwidth]{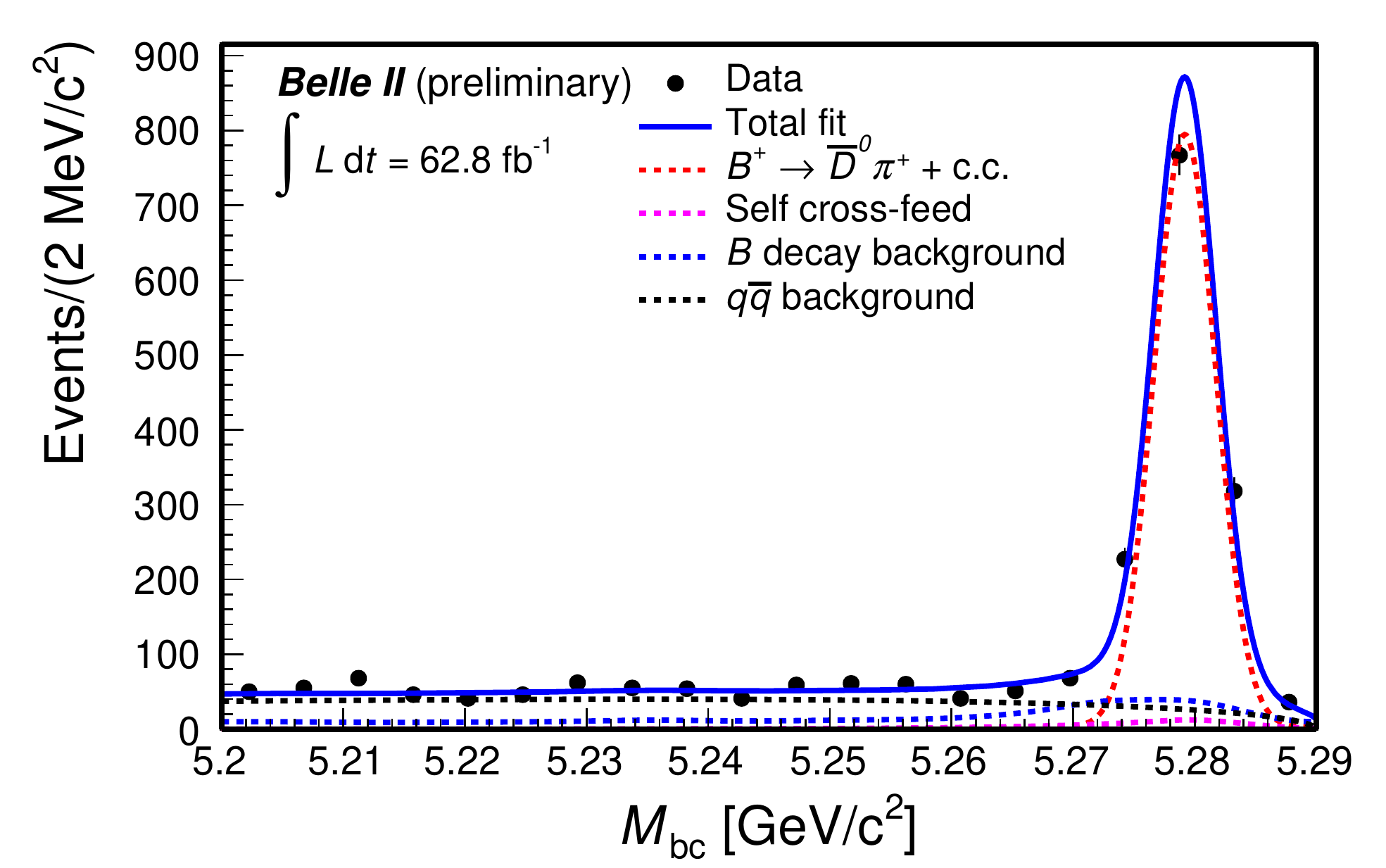}
\includegraphics[width=0.475\textwidth]{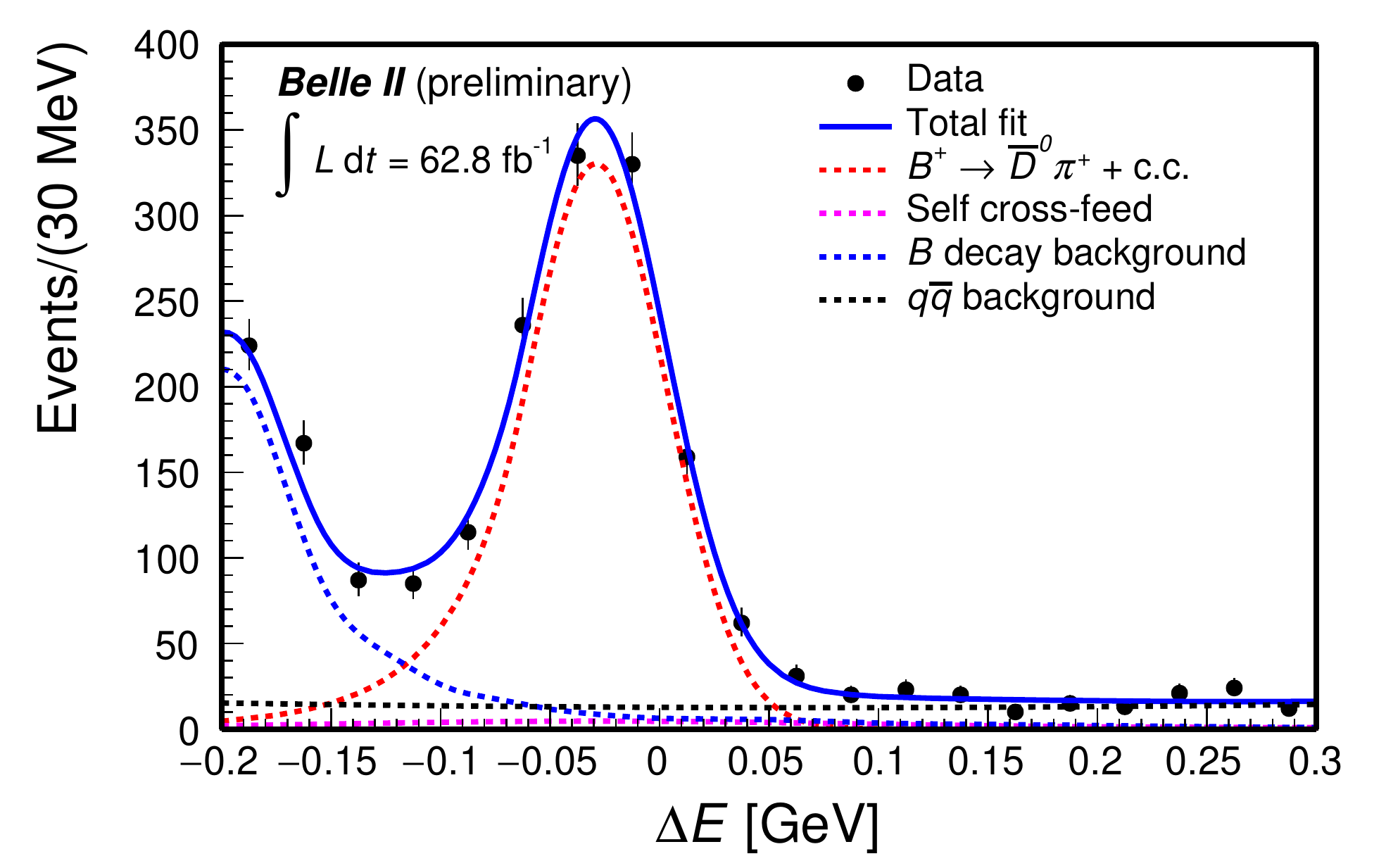}
\caption{Distributions of (left) $M_{bc}$ and (right) $\Delta E$ for $B^+\to \overline{D}^0 (\to K^+\pi^-\pi^0)\pi^+$ candidates reconstructed in (top) simulated events and (bottom) data selected with an optimized continuum-suppression requirement, and
projected onto the signal region (left panel: $-0.14 < \Delta E < 0.06$ GeV, right panel: $M_{bc} > 5.27$ $\text{GeV}/c^2$). The projections of the fit are overlaid.}
\label{fig:CR_fitting_results}
\end{figure}

The resulting difference in the means of $\Delta E$ between data and simulation is $-17.0 \pm 1.4$ MeV, and the resulting width scale-factor is about $8\%$ larger in data. We use these corrections in the fit to the signal sample. No data-simulation discrepancy has been observed for the $M_{bc}$ distribution. Figure~\ref{fig:CR_fitting_results} shows the $M_{bc}$-$\Delta E$ distributions with fit projections overlaid, in the signal region, $M_{bc} > 5.27$ GeV$/c^2$ and $-0.14 < \Delta E < 0.06$ GeV.

\label{sec:Consis_check}

%% file: contents/Determination_of_Signal_Yields.tex
\section{Signal Determination}
\label{sec:Yeilds}
The total signal yields and {\it CP}-violating charge-asymmetries are obtained using extended maximum-likelihood fits of the $M_{bc}$ and $\Delta E$ distributions.
The probability density function (PDF) for $\Delta E$ is modeled by a Crystal Ball function and a Gaussian function with a common peak position to describe the long tail in the low $\Delta E$ region due to photon-shower leakage in the calorimeter. The $M_{bc}$ PDF is described by two Gaussian functions with a common mean. All the PDF parameters are first obtained using the Monte Carlo simulation and then corrected for MC-data discrepancies based on the control sample study described in Section \ref{sec:Consis_check}.  
The PDF for the $B\overline{B}$ background is modeled by a two-dimension kernel function, which is a superposition of Gaussian functions for each $M_{bc}$-$\Delta E$ point. The PDF for the continuum background is modeled as the product of an Argus function in $M_{bc}$ and a second order polynomial in $\Delta E$.

The likelihood function for each mode is
\begin{equation}
\begin{split}
    \textit{L} = \frac{e^{-\sum\nolimits_{i}N_{j}}}{N!}\prod_{i=1}^{N}(\sum\nolimits_{i}N_{j}P_{j}^i(M_{bc},\Delta E))~,~\text{where}\\
    P_{j}^i(M_{bc},\Delta E) = \frac{1}{2}(1-q_{i}\cdot\mathcal{A}_{\it raw})\times P_{j}(M_{bc},\Delta E)~.
\end{split}
\label{eqn:ML}
\end{equation}
Here $i$ is the event index and $N_j$ is the yield of events for the category $j$, which indexes signal, continuum, feed-across, and other $B$ decays. Here $P_{j}^i(M_{bc},\Delta E)$ is the probability density function for the $i$th event, $q_{i}$ denotes the candidate’s flavor (charge) ($q_{i}$ = +1 for $B^+$ and $q_{i}$ = $-1$ for $B^-$), and the yield asymmetry is defined as
\begin{equation}
    \mathcal{A}_{\it raw} = \frac{N_{B^-}-N_{B^+}}{N_{B^-}+N_{B^+}}~,
\label{eqn:Acp}
\end{equation}
where $N_{B^+ (B^-)}$ is the yield of $B^+ (B^-)$ candidates.
\\\indent
We simultaneously fit for $B^+ \to K^+\pi^0$ and $\pi^\pm\pi^0$ yields, which feed-across to each other, to determine the branching fractions and {\it CP}-violating asymmetries. The fraction of signal and feed-across candidates are constrained with Gaussians according to the mean values and statistical uncertainties of corrected ratios ($\varepsilon_{\rm MC}$/$\varepsilon_{\rm Data}$), which are about $96\%$, where $\varepsilon_{\rm MC}$ and $\varepsilon_{\rm Data}$ are PID efficiencies or fake rates in simulation and data, respectively. 

To deal with the simulation-data discrepancies in PDF modeling, we fix the shapes of the signals and feed-across, but float their mean shifts and width ratios in $M_{bc}$ and $\Delta E$. The range is constrained to be a Gaussian, whose mean and width is determined from the mean shift and the multiplicative scale-factor in the control sample study. For the $B\overline{B}$ background, we fix the shape and shift the model in $\Delta E$ with the mean shift that we obtained from the control sample study. For the $q\overline{q}$ background, we float the shape parameters.

%% file: contents/Measurements_for_Branching_Factions_and_CP_Asymmetry.tex
\section{Measurements of Branching Fractions and {\it CP} Asymmetry}
\label{sec:Results}
\begin{figure}[htb]
\centering
\includegraphics[width=0.475\textwidth]{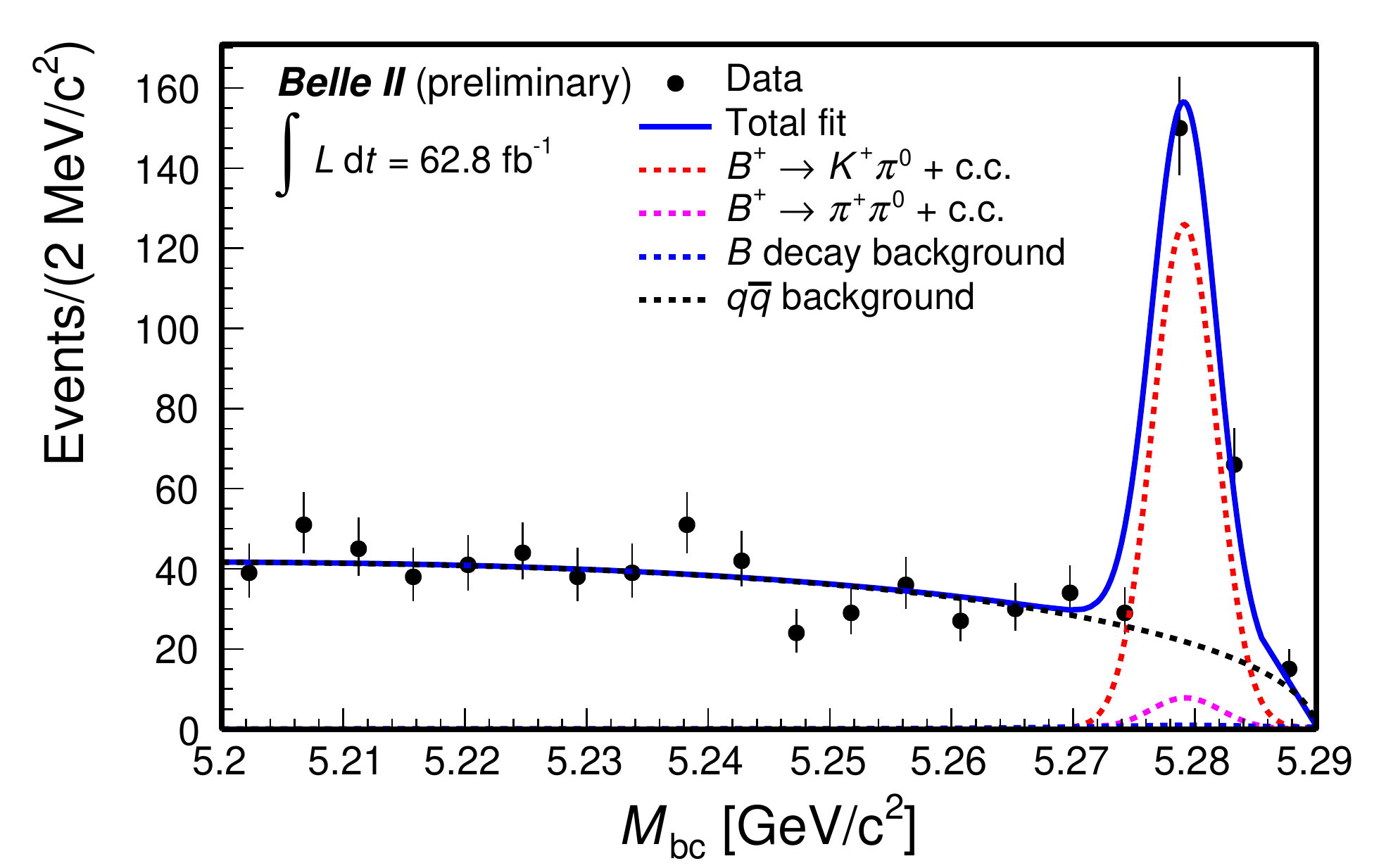}
\includegraphics[width=0.475\textwidth]{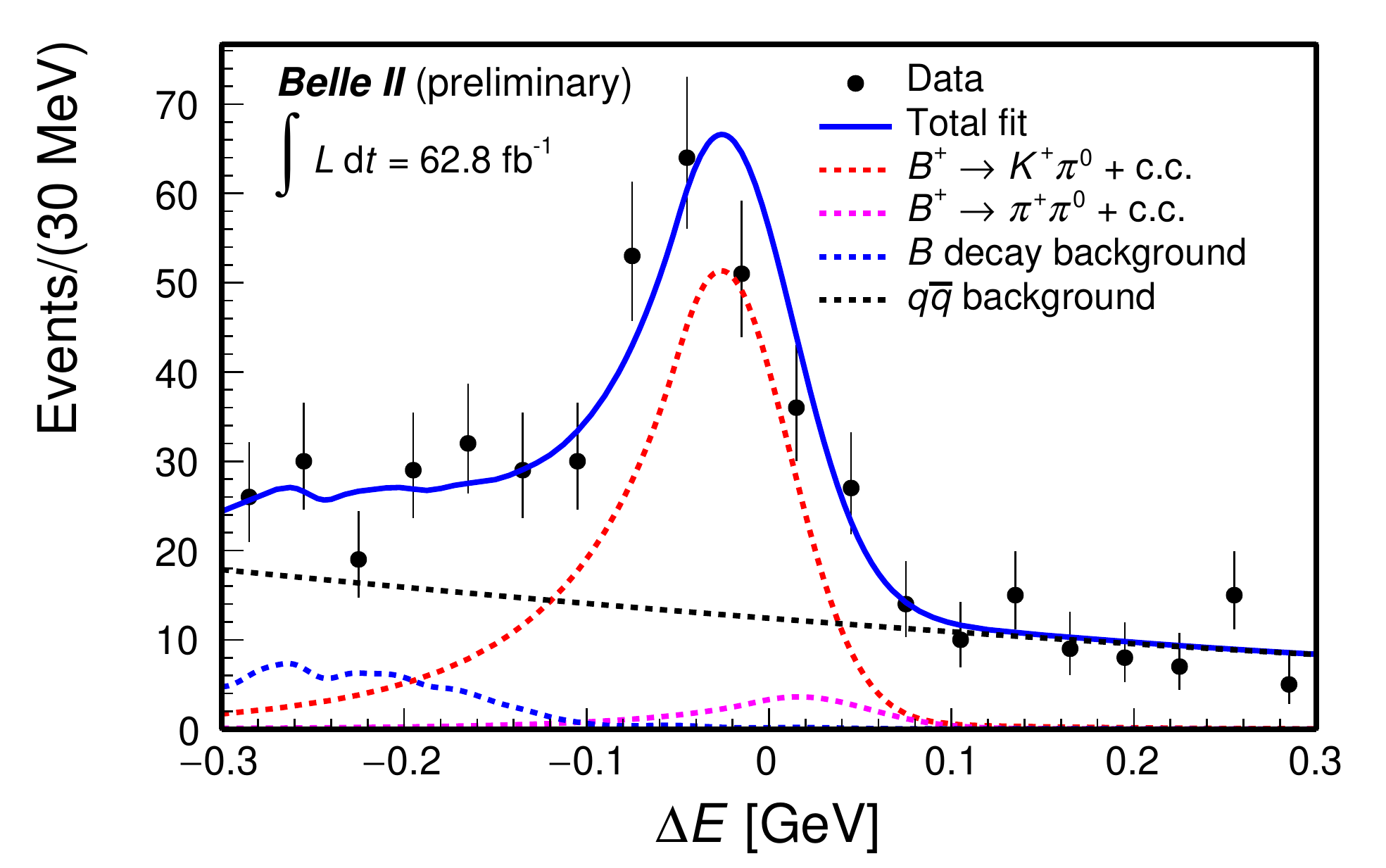}\\
{a) $B^+\to K^+\pi^0$}\vspace{2mm}\\
\includegraphics[width=0.475\textwidth]{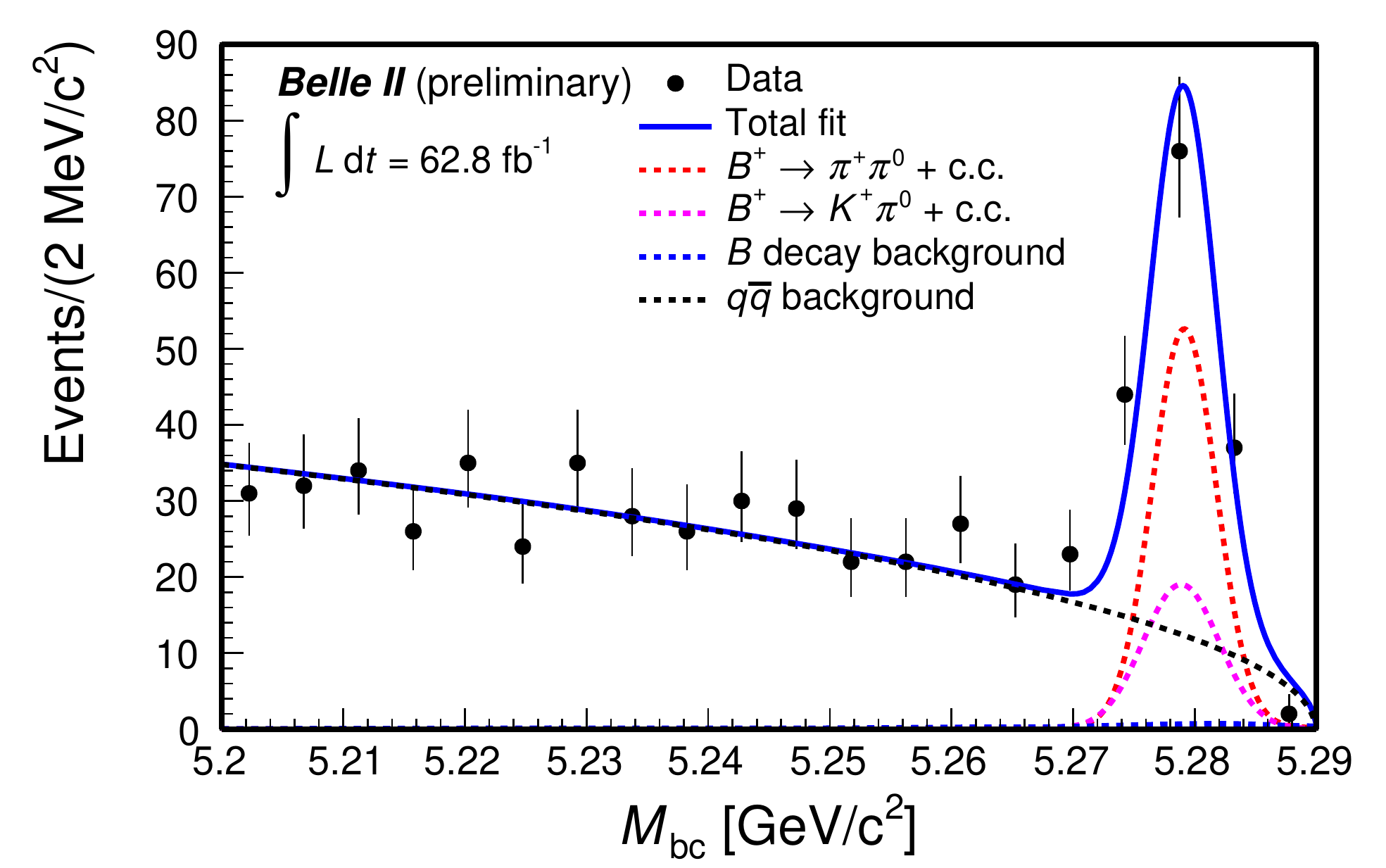}
\includegraphics[width=0.475\textwidth]{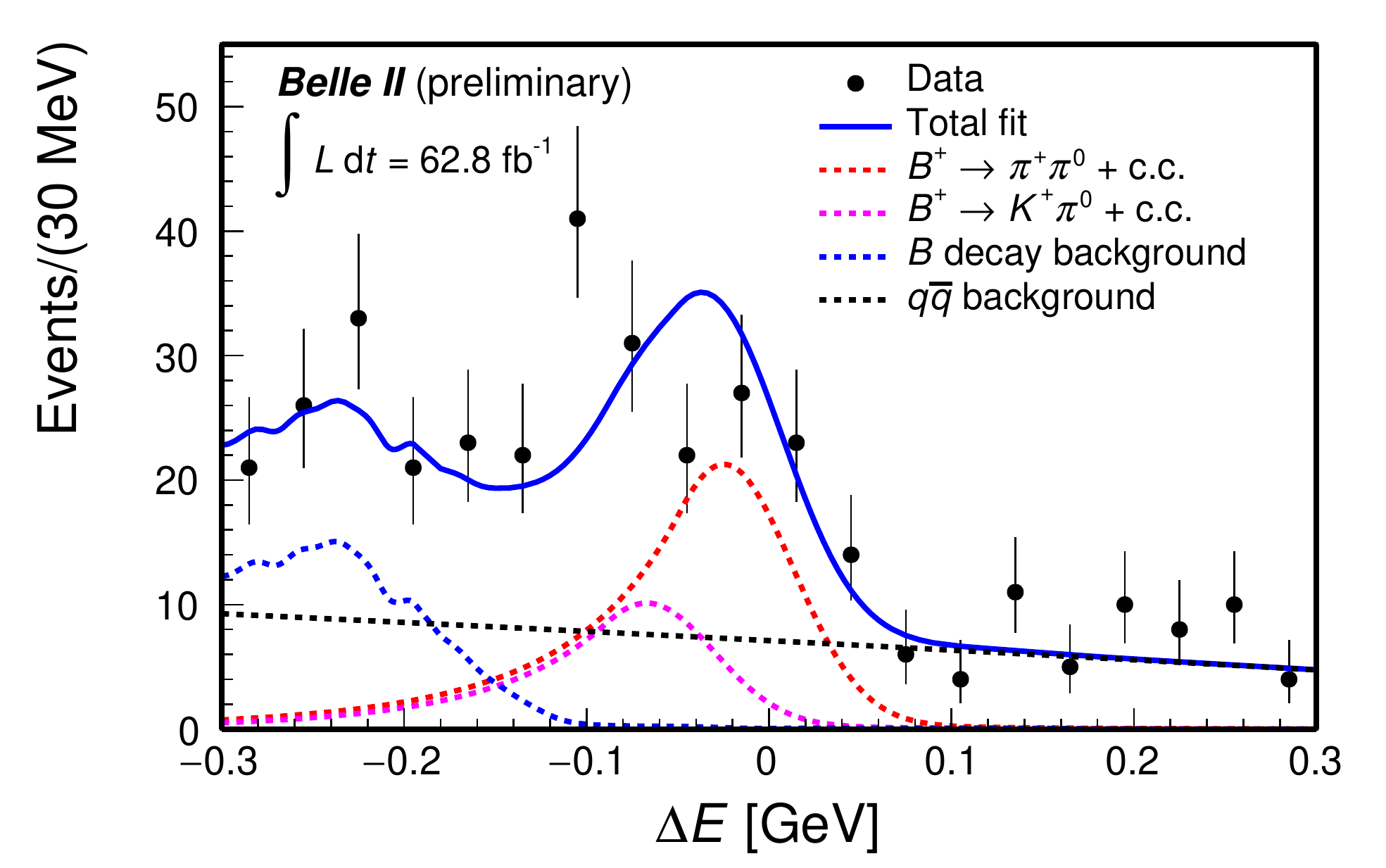}\\
{b) $B^+\to \pi^+\pi^0$}\vspace{2mm}\\
\caption{Distributions of (left) $M_{bc}$ and (right) $\Delta E$ for $B^+ \to h^+ \pi^0$ candidates reconstructed in 2019–2020 Belle~II data selected with an optimized continuum-suppression requirement, and projected onto the signal region (left panel: $-0.14 < \Delta E < 0.06$ GeV, right panel: $M_{bc} > 5.27$ $\text{GeV}/c^2$). The projections of the fit are overlaid.}
\label{fig:fitting_rs_data_moriond}
\end{figure}

\begin{figure}[htb]
\centering
\includegraphics[width=0.475\textwidth]{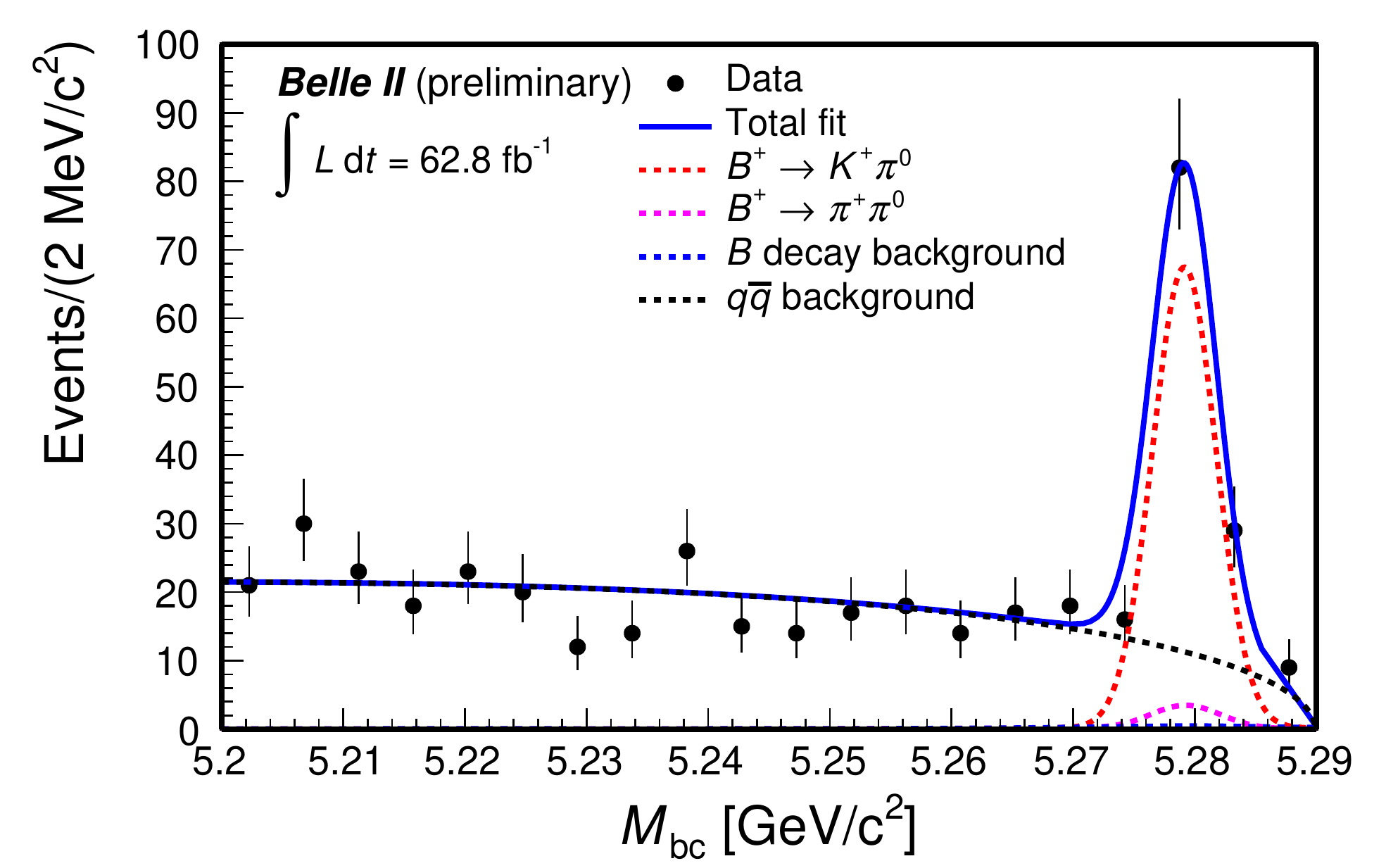}
\includegraphics[width=0.475\textwidth]{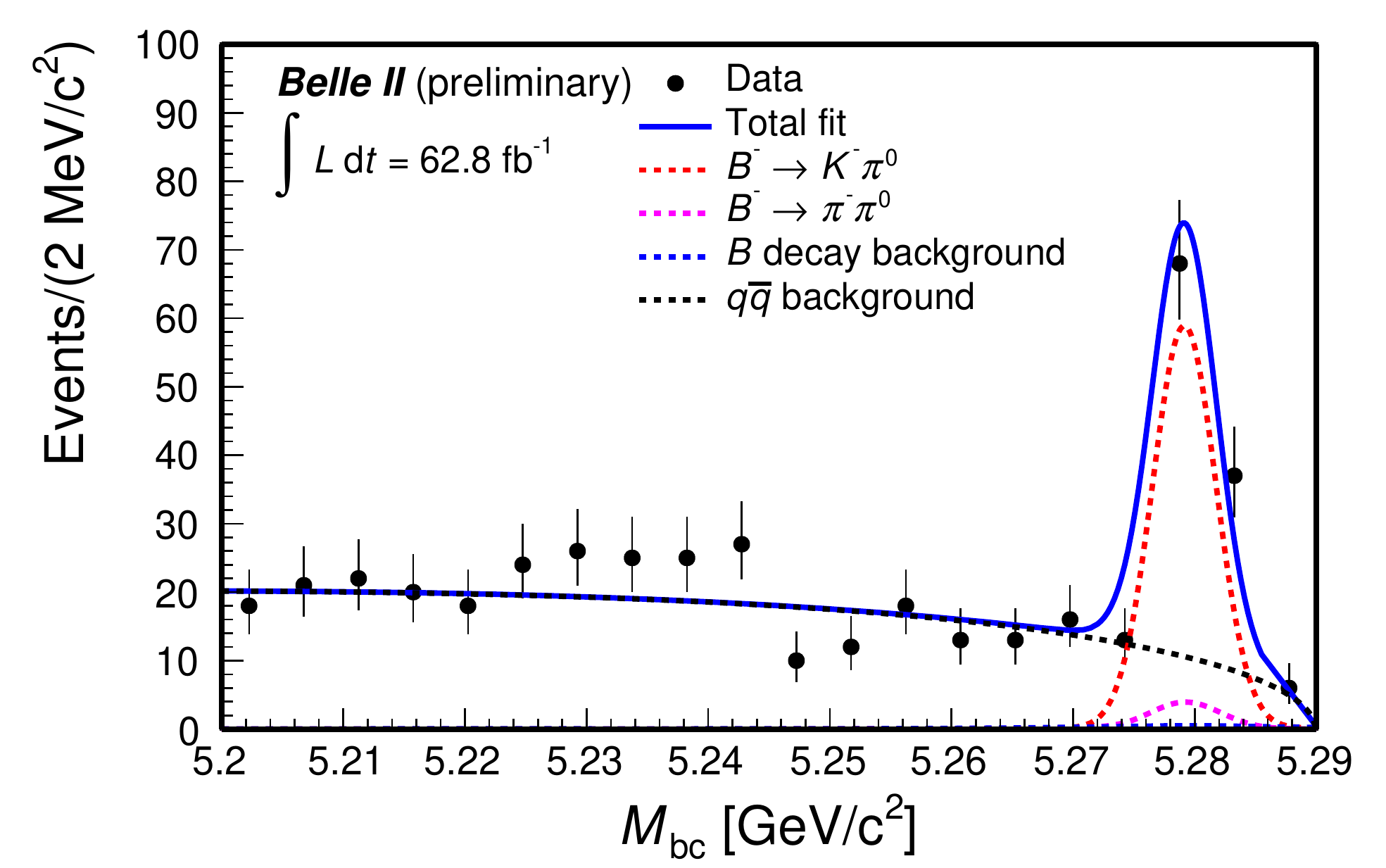}
\includegraphics[width=0.475\textwidth]{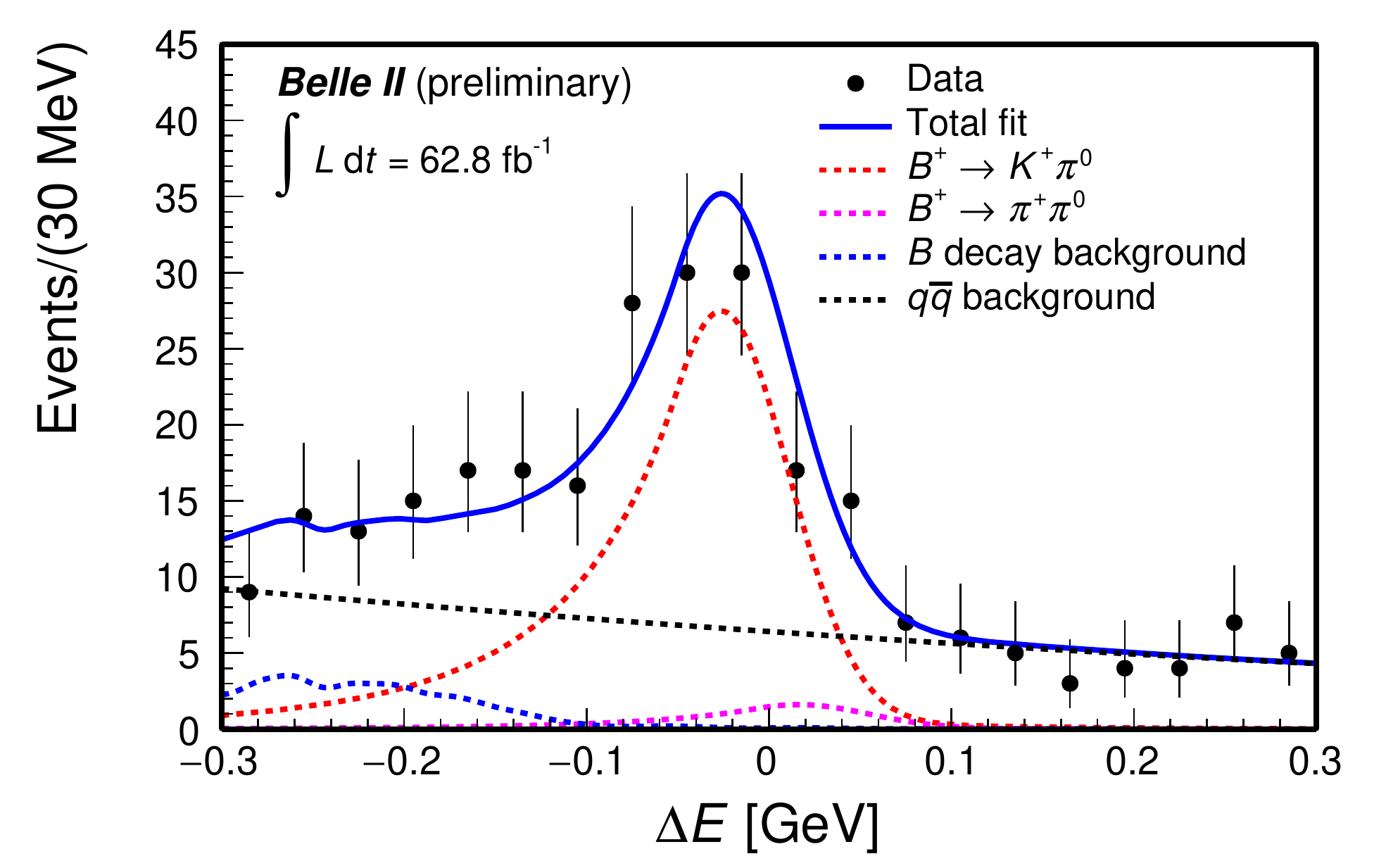}
\includegraphics[width=0.475\textwidth]{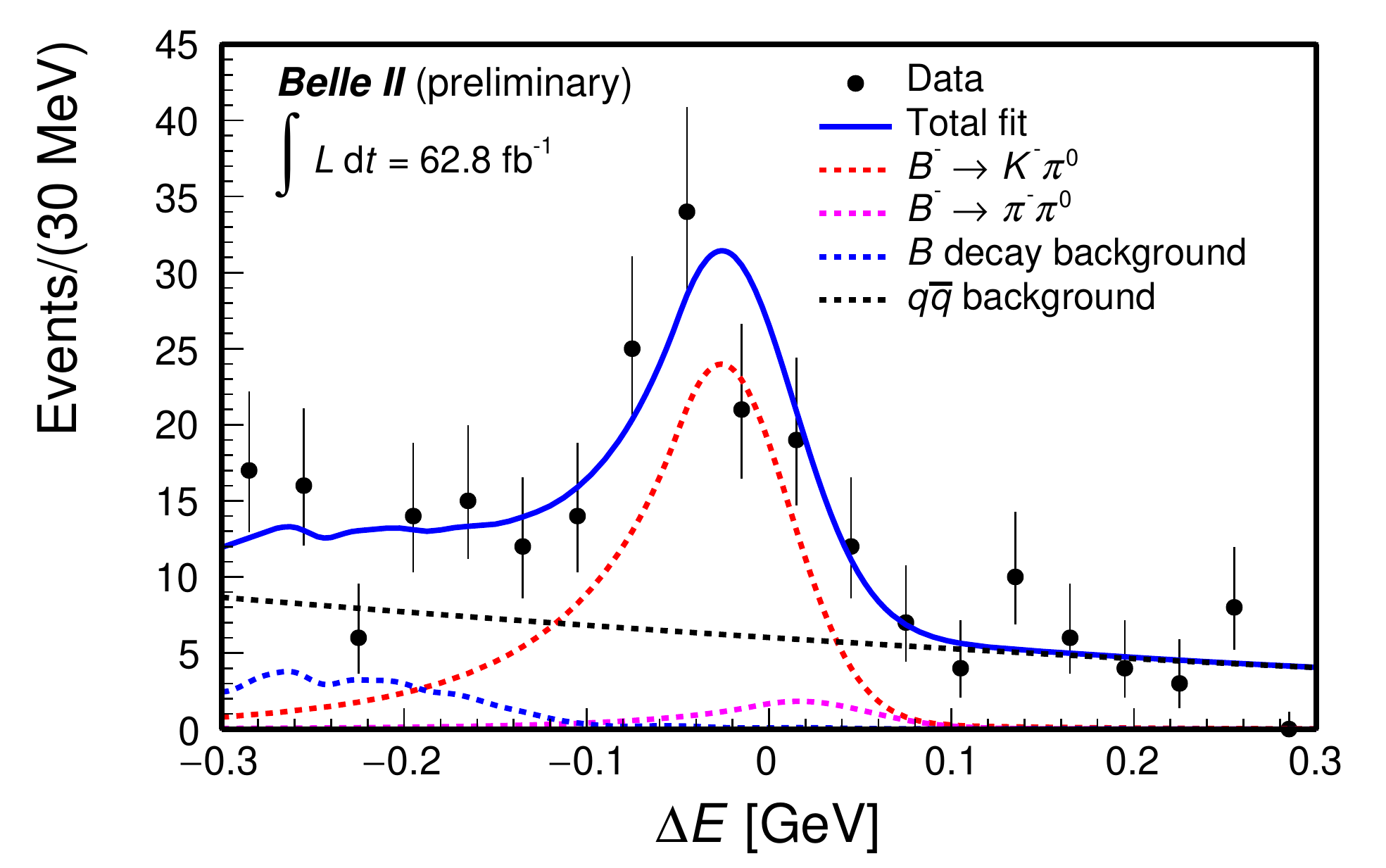}
\caption{Charge-specific distributions of (top) $M_{bc}$ and (bottom) $\Delta E$ for (left) $B^+ \to K^+ \pi^0$ and (right) $B^- \to K^- \pi^0$ candidates reconstructed in 2019–2020 Belle~II data selected with an optimized continuum-suppression requirement, and projected onto the signal region (top panel: $-0.14 < \Delta E < 0.06$ GeV, bottom panel: $M_{bc} > 5.27$ $\text{GeV}/c^2$). The projections of the fit are overlaid.}
\label{fig:fitting_rs_Kpi_data_acp_moriond}
\end{figure}

\begin{figure}[htb]
\centering
\includegraphics[width=0.475\textwidth]{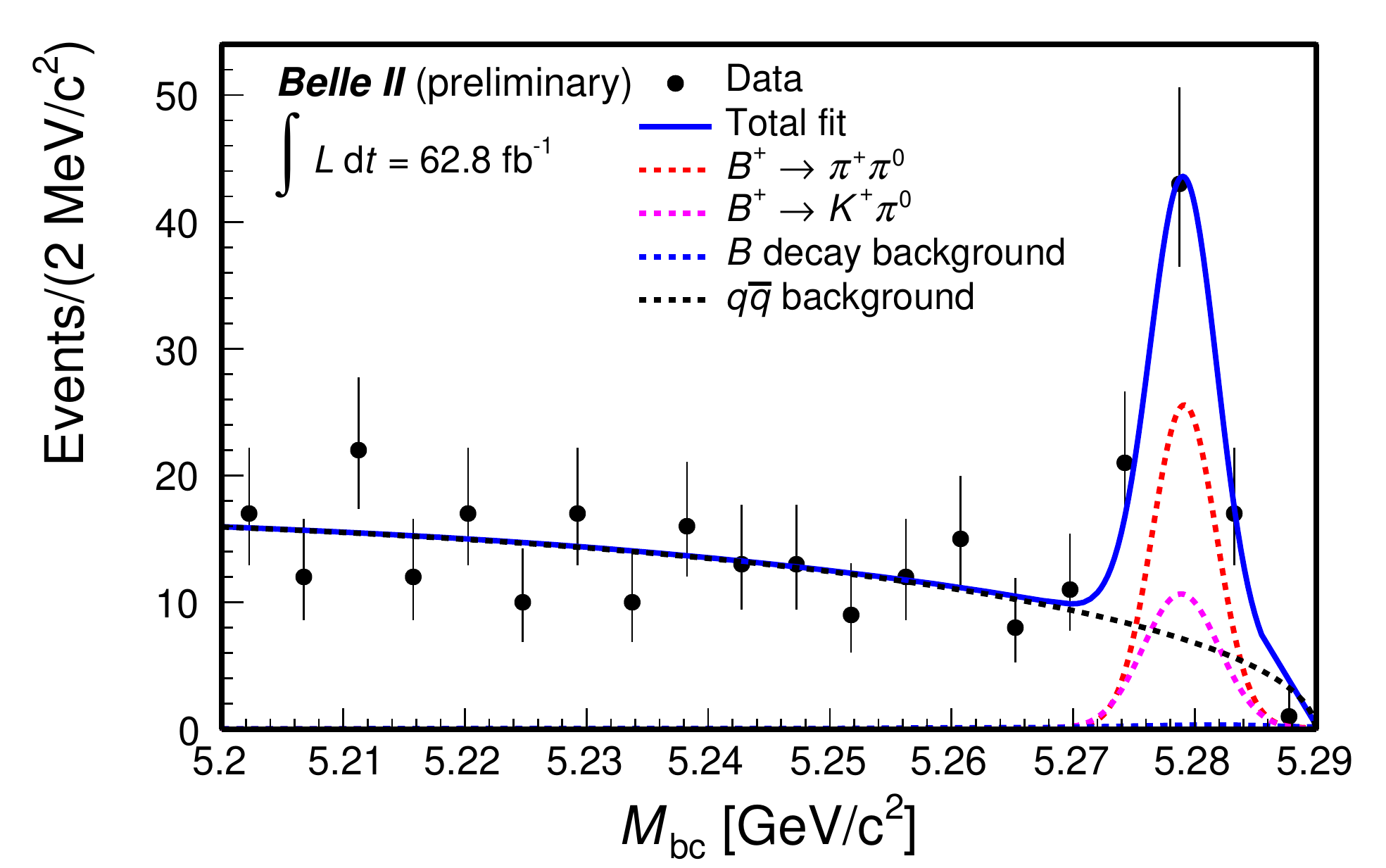}
\includegraphics[width=0.475\textwidth]{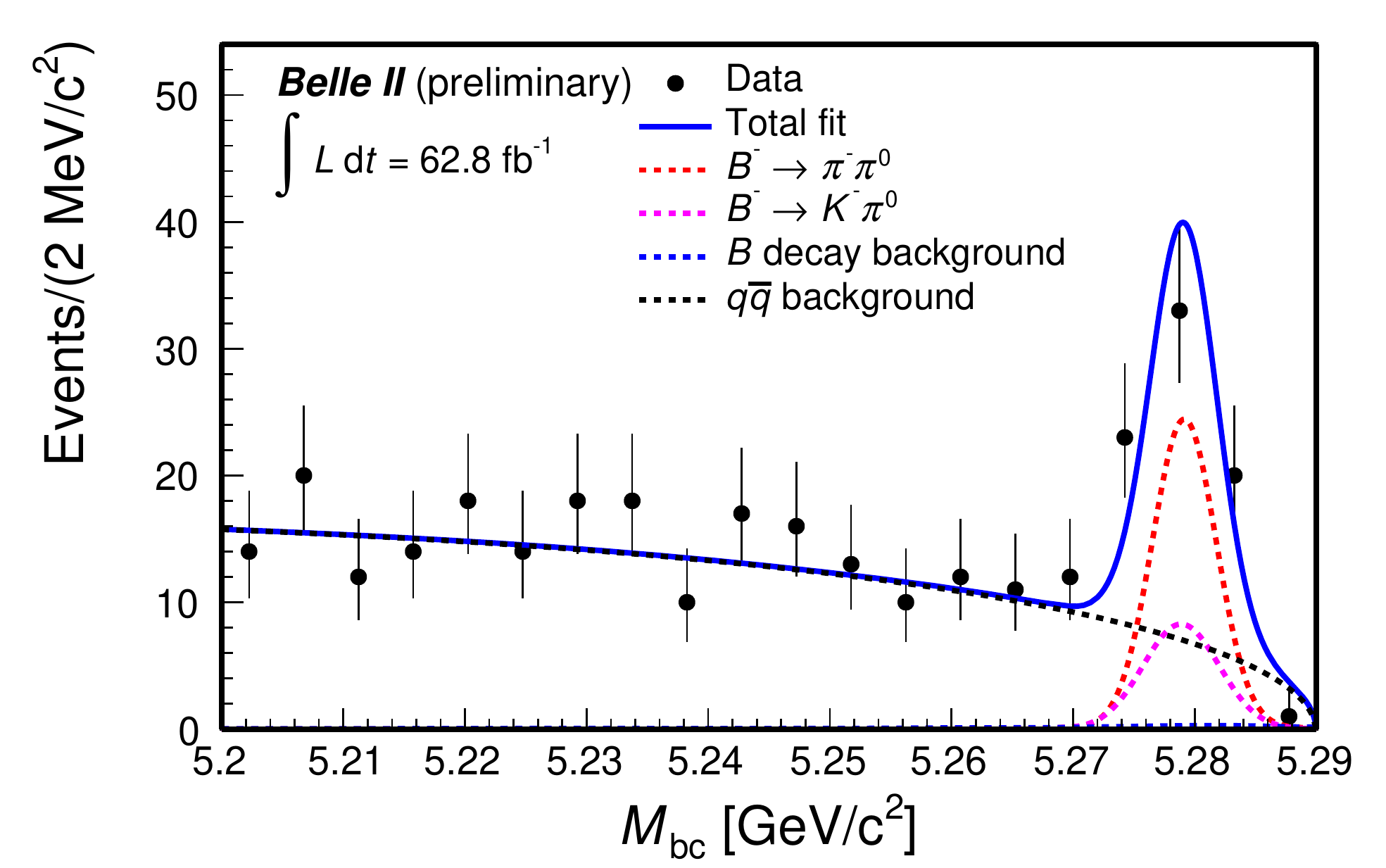}
\includegraphics[width=0.475\textwidth]{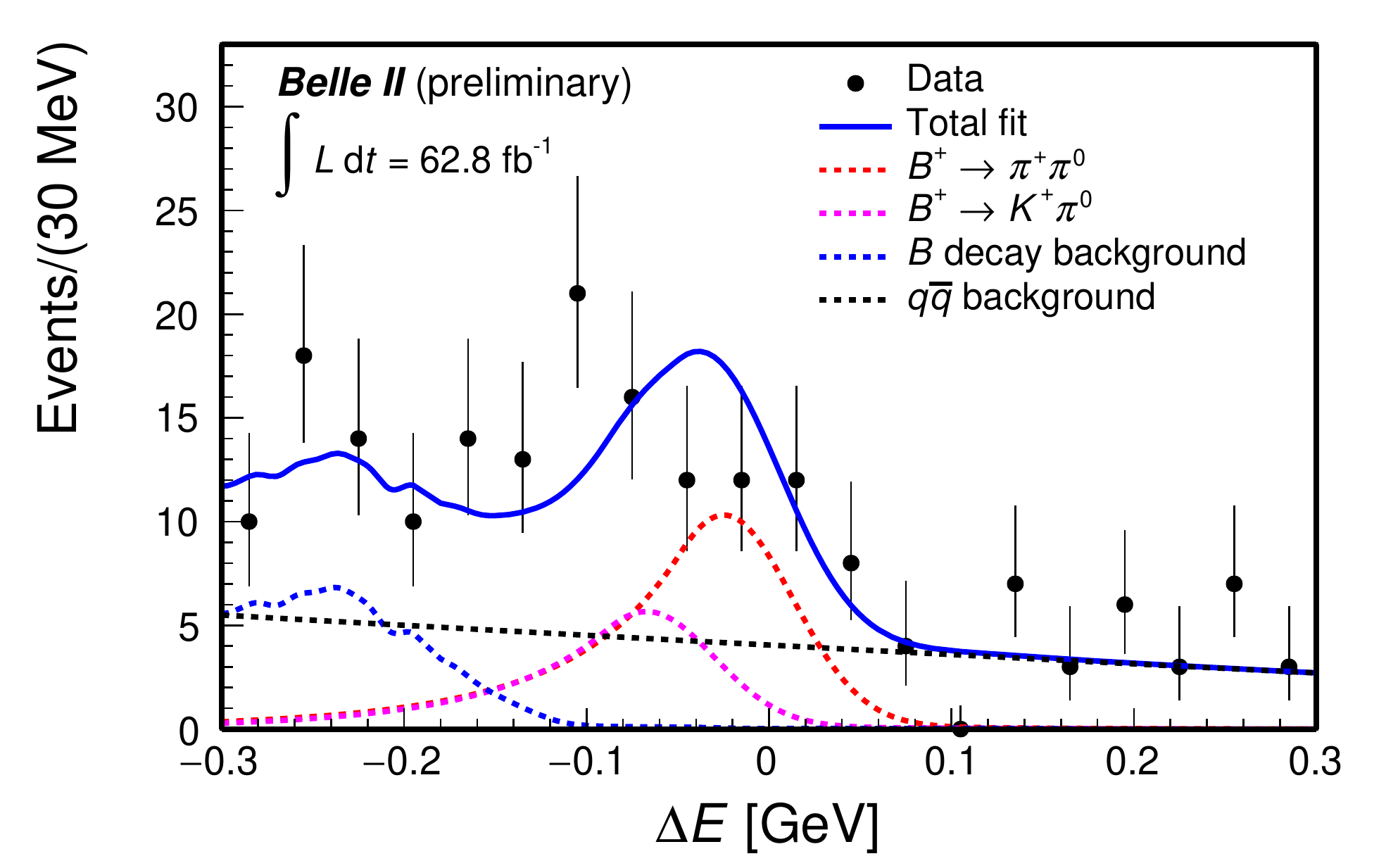}
\includegraphics[width=0.475\textwidth]{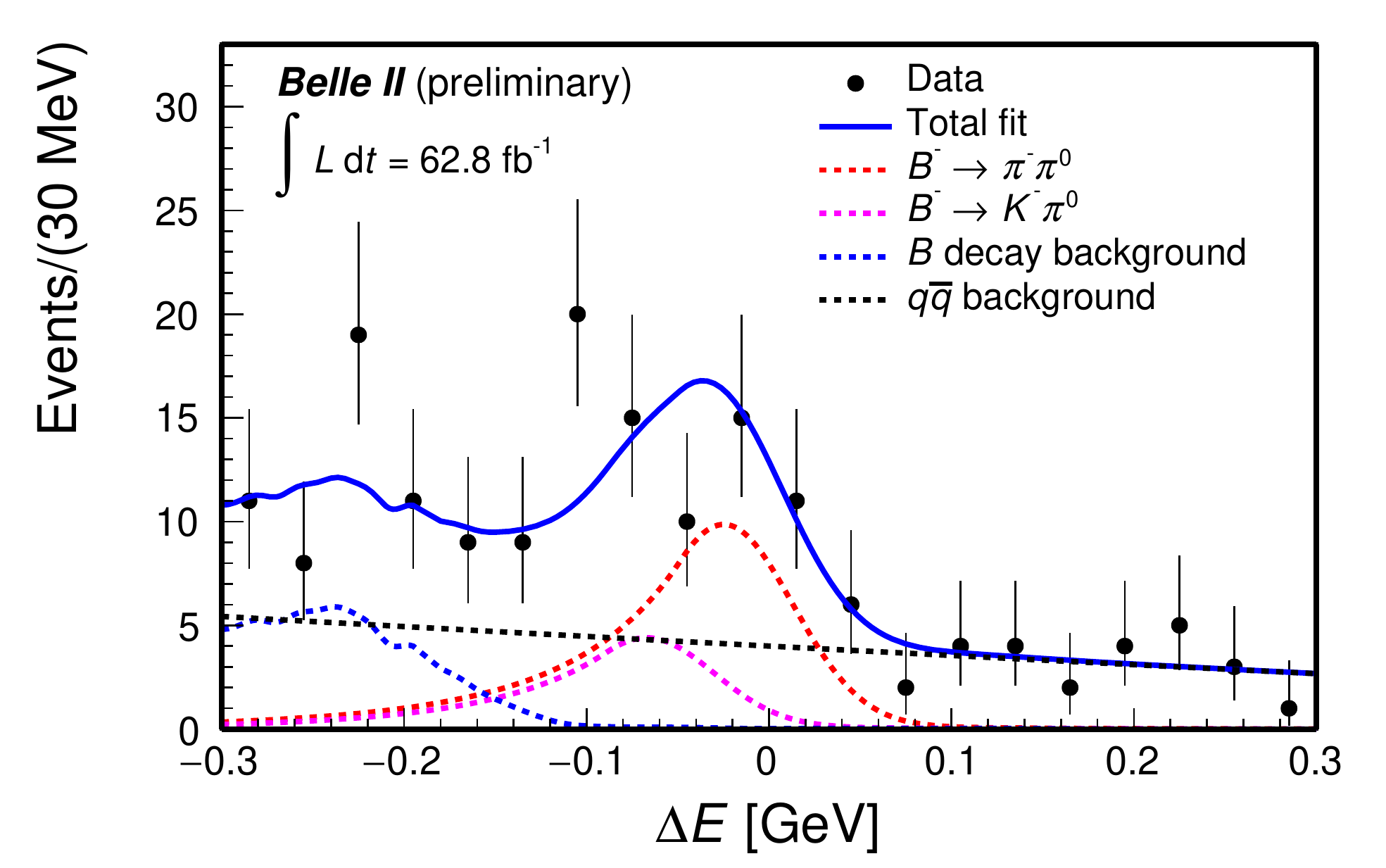}
\caption{Charge-specific distributions of (top) $M_{bc}$ and (bottom) $\Delta E$ for (left) $B^+ \to \pi^+ \pi^0$ and (right) $B^- \to \pi^- \pi^0$ candidates reconstructed in 2019–2020 Belle~II data selected with an optimized continuum-suppression requirement, and projected onto the signal region (top panel: $-0.14 < \Delta E < 0.06$ GeV, bottom panel: $M_{bc} > 5.27$ $\text{GeV}/c^2$). The projections of the fit are overlaid.}
\label{fig:fitting_rs_pipi_data_acp_moriond}
\end{figure}

The $M_{bc}$-$\Delta E$ distributions of $B^+ \to h^+ \pi^0$ data are shown in Figure~\ref{fig:fitting_rs_data_moriond}, with the fit curve overlaid. Figures~\ref{fig:fitting_rs_Kpi_data_acp_moriond}, \ref{fig:fitting_rs_pipi_data_acp_moriond} show the fits on charge-specific events overlaid on the projection of the ($M_{bc}$, $\Delta E$) dimensions. The details of fit results are given below.
\subsection{Branching fractions}
The branching fractions are determined as  
\begin{equation}
\begin{aligned}
    \mathcal{B}_{B^\pm \to h^\pm \pi^0} = \frac{N_{B^+ \to h^+ \pi^0}}{\epsilon_{h^+ \pi^0} \times \mathcal{R}^+_{\rm PID} \times (1-\mathcal{A}_{\it raw}) \times N_{B\overline{B}}} + \frac{N_{B^- \to h^- \pi^0}}{\epsilon_{h^- \pi^0} \times \mathcal{R}^-_{\rm PID} \times (1+\mathcal{A}_{\it raw}) \times N_{B\overline{B}}}~.
\label{eqn:BR_est}
\end{aligned}
\end{equation}
Here $\epsilon_{h^\pm \pi^0}$ is the efficiency for selecting and reconstructing signal as determined from simulation, $\mathcal{R}^\pm_{\rm PID}$ is the ratio of PID selection efficiencies in data and simulation for positive and negative charged particles, $N_{B^+ \to h^+ \pi^0}$ is the number of generated events corresponding to the data luminosity, and $N_{B\overline{B}}$ is the number of $B\overline{B}$ pairs in the data sample, corresponding to 35.8 million for $B^+B^-$. We obtain the number of ${\PB\overline{B}}$~pairs from the measured integrated luminosity, the \mbox{$\Pep\Pem\to\Upsilon(4{\rm S})$}~cross section~$(1.110 \pm 0.008)\,$nb~\cite{Bevan:2014iga}~(assuming that the $\Upsilon(4{\rm S})$ decays exclusively to ${\PB\overline{B}}$~pairs), and the \mbox{$\Upsilon(4{\rm S})\to\PBzero\overline{B}^0$}~branching fraction \mbox{$f^{00} = 0.487\pm 0.010\pm 0.008$}~\cite{Aubert:2005bq}. Table~\ref{tab:BR_detail_moriond} summarizes the result of fit for branching fractions, signal yields and corrected signal efficiencies ($\epsilon \times \mathcal{R}_{\rm PID}$).
\begin{table}[!ht]
\begin{center}
\caption{Fit results for $B^+ \to h^+ \pi^0$ branching fractions. The product $\epsilon \times \mathcal{R}_{\rm PID}$ is the corrected signal efficiency.}
\begin{tabular}{l c c}
\hline \hline
 & $B^+ \to K^+ \pi^0$ & $B^+ \to \pi^+ \pi^0$\\
\hline 
$\mathcal{B}\ [10^{-6}]$\ & $11.9 ^{+1.1}_{-1.0}$ & $5.5 ^{+1.0}_{-0.9}$\\
Signal yields & $211.0 ^{+18.8}_{-18.0}$ & $83.9 ^{+14.7}_{-13.9}$\\
$\epsilon \times \mathcal{R}_{\rm PID}$ & $0.247 \pm 0.003$ & $0.212 \pm 0.002$\\
Mean shift (MeV/$c^2$) in $M_{bc}$ & \multicolumn{2}{c}{$-0.07 \pm 0.09$}\\
Scale factor in $M_{bc}$ & \multicolumn{2}{c}{$0.99 \pm 0.03$}\\
Mean shift (MeV) in $\Delta E$ & \multicolumn{2}{c}{$-17.63 ^{+1.35}_{-1.36}$}\\
Scale factor in $\Delta E$ & \multicolumn{2}{c}{$1.11 \pm 0.04$}\\
\hline \hline
\end{tabular} 
\label{tab:BR_detail_moriond}
\end{center}
\end{table}
\label{subsec:BR}
\subsection{{\it CP}-violating asymmetry}
In general, $\mathcal{A}_{\it raw}$ is due to the combination of genuine {\it CP}-violating effects ($\mathcal{A}_{\it CP}$) in the
decay dynamics and instrumental asymmetries due to differences in interaction or reconstruction probabilities between opposite-charge hadrons. Such a combination is additive for small asymmetries, $\mathcal{A}_{\it raw} = \mathcal{A}_{\it CP} + \mathcal{A}_{det}$, where $\mathcal{A}_{det}$ is the instrumental asymmetry measured by reconstructing the decays $\overline{D}^+ \to K^0_S \pi^+$ and $\overline{D}^0 \to K^+ \pi^-$ to estimate $\mathcal{A}(\pi)$ and $\mathcal{A}(K)$ by $\mathcal{A}(\pi)$ = $\mathcal{A}(K_{S}^0\pi) - \mathcal{A}(K_{S}^0)$, $\ \mathcal{A}(K)$ = $\mathcal{A}(K\pi) - \mathcal{A}(K_{S}^0\pi) + \mathcal{A}(K_{S}^0)$. Table~\ref{tab:ins_acp} shows the results for instrumental asymmetries.
\begin{table}[!ht]
\begin{center}
\caption{Instrumental charge-asymmetries associated with $K^\pm\pi^\mp$, $K_{S}^0\pi^\pm$,  $K^\pm$, and $\pi^\pm$  reconstruction, obtained using samples of $D^0 \to K^- \pi^+$ and $D^+ \to K_{S}^0 \pi^+$ decays.}
\begin{tabular}{l  c  c}
\hline\hline
Instrumental asymmetry & Value \\
\hline
$\mathcal{A}_{det}(K^+\pi^-)$   & $-0.010 \pm 0.001$ \\
$\mathcal{A}_{det}(K_{S}^0\pi^+)$   & $+0.026 \pm 0.019$ \\
$\mathcal{A}_{det}(K^+)$   & $+0.017 \pm 0.019$ \\
$\mathcal{A}_{det}(\pi^+)$   & $+0.026 \pm 0.019$ \\
\hline\hline
\end{tabular}
\label{tab:ins_acp}
\end{center}
\end{table}
Table~\ref{tab:acp_detail_moriond} summarizes the results for $\mathcal{A}_{\it CP}$, $\mathcal{A}_{\it raw}$, signal yields, and corrected signal efficiencies ($\epsilon \times \mathcal{R}^+_{\rm PID}$ and $\epsilon \times \mathcal{R}^-_{\rm PID}$) for $B^+$ and $B^-$ decays. 
\begin{table}[!ht]
\begin{center}
\caption{Fit results for $B^+ \to h^+ \pi^0$ {\it CP}-violating asymmetries. The product $\epsilon  \times \mathcal{R}^\pm_{\rm PID}$ is the corrected signal efficiency for positively (negatively) charged $h^+$ candidates.}
\begin{tabular}{l c c}
\hline \hline
 & $B^+ \to K^+ \pi^0$ & $B^+ \to \pi^+ \pi^0$\\
\hline 
$\mathcal{A}_{\it CP}$ & $-0.089 \pm 0.085$ & $-0.042 \pm 0.166$\\
$\mathcal{A}_{\it raw}$ & $-0.072 \pm 0.085$ & $-0.016 \pm 0.166$\\
$B^+$ yields & $111.1 \pm 13.0$ & $42.2 \pm 9.8$\\
$\epsilon \times \mathcal{R}^+_{\rm PID}$ & $0.243 \pm 0.003$ & $0.210 \pm 0.002$\\
$B^-$ yields & $99.8 \pm 12.5$ & $41.7 ^{+9.8}_{-9.9} $\\
$\epsilon \times \mathcal{R}^-_{\rm PID}$ & $0.252 \pm 0.003$ & $0.214 \pm 0.002$\\
\hline \hline
\end{tabular} 
\label{tab:acp_detail_moriond}
\end{center}
\end{table}

\label{subsec:ACP}

%% file: contents/Systematic_Error_Extra.tex
\section{Systematic Uncertainties}
\label{sec:sys}
We consider several sources of systematic uncertainties. We assume the sources to be independent, and add in quadrature the corresponding uncertainties. Table~\ref{tab:tol_BR_sys} and~\ref{tab:tol_acp_sys} summarize systematic uncertainties for the measurements of the branching fractions and {\it CP}-violating asymmetries, respectively.

The systematic uncertainty due to data-simulation discrepancies in charged-particle reconstruction efficiency~\cite{Tracking:2020} is estimated to be $0.91\%$ per track. The systematic uncertainty on $\pi^0$ reconstruction is studied using the decays $B^0 \to D^{*-}(\to \overline{D}^0 (\to K^+\pi^-)\pi^-) \pi^+$, $B^0 \to D^{*-}(\to \overline{D}^0 (\to K^+\pi^-\pi^0)\pi^-) \pi^+$ and the ratio of yields between data and simulation. Because the efficiency ratio is compatible with unity, the statistical uncertainty of the ratio is quoted as the systematic uncertainty.
The systematic uncertainties in continuum suppression efficiency are evaluated from data-MC discrepancies in the continuum-suppression efficiency using the control channel $B^+ \to \overline{D}^0(\to K^+\pi^-\pi^0) \pi^+$. The selection efficiencies obtained in data and simulation agree well; hence the statistical uncertainties on the ratio of data to MC efficiencies are assigned as systematic uncertainties. The systematic uncertainty on the number of $B\overline{B}$ pairs is $1.4\%$~\cite{Abudinen:2019osb}.  

The fitting systematic uncertainties are due to PDF mismodeling of each component. The uncertainties for signal, feed-across, and continuum are estimated from the difference between the average of results of the sample-composition fit performed on ensembles of simplified simulated experiments generated with the baseline and alternate models. The systematic uncertainties due to data-MC discrepancies in the $B\overline{B}$ background distributions are evaluated by measuring the difference in branching fractions with a more restrictive fitting region, $\Delta E > -0.12~(\text{GeV})$.  

The systematic uncertainties for the {\it CP}-violating asymmetry results due to PDF mismodeling of signal, feed-across, and continuum are estimated from the difference in {\it CP}-violating asymmetries while performing fits with an alternative model for each component. The model for each component is assumed to be charge symmetric. We estimate the systematic uncertainties due to model asymmetry by measuring the difference in {\it CP}-violating asymmetry with independent charge-specific continuum shape parameters and signal, feed-across mean shifts, and width ratios in $M_{bc}$ and $\Delta E$. The systematic uncertainties due to data-MC discrepancies in the $B\overline{B}$ background distributions are estimated with the same procedure as applied for the branching fractions. A systematic error associated with the uncertainty on the instrumental asymmetry (Table~\ref{tab:ins_acp}) is also included.
\begin{table}[!ht]
\begin{center}
\caption{Systematic uncertainties in the branching fraction measurements. The total systematic uncertainty is the sum in quadrature of all the terms.}
\begin{tabular}{l  c  c}
\hline\hline
Source & $B^+ \to K^+ \pi^0$ & $B^+ \to \pi^+ \pi^0$ \\
\hline
Tracking & $0.91\%$ & $0.91\%$ \\
$\pi^0$ efficiency & $13.00\%$ & $13.00\%$ \\
Continuum suppression & $2.90\%$ & $3.30\%$ \\
$N_{B\overline{B}}$ & $1.40\%$ & $1.40\%$ \\
Signal, feed-across model & $0.84\%$ & $1.37\%$ \\
$B\overline{B}$ background model & $0.40\%$ & $1.44\%$ \\
$q\overline{q}$ background model & $0.55\%$ & $2.49\%$ \\
\hline
Total & $13.47\%$ & $13.89\%$ \\
\hline\hline
\end{tabular}
\label{tab:tol_BR_sys}
\end{center}
\end{table}
\begin{table}[!ht]
\begin{center}
\caption{Systematic uncertainties in the {\it CP}-violating asymmetry measurements. The total systematic uncertainty is the sum in quadrature of all the terms.}
\begin{tabular}{l  c  c}
\hline\hline
Source & $B^+ \to K^+ \pi^0$ & $B^+ \to \pi^+ \pi^0$ \\
\hline
Detector bias & $0.019$ & $0.019$ \\
Signal, feed-across model & $0.008$ & $0.004$ \\
Background model & $0.014$ & $0.059$ \\
\hline
Total & $0.025$ & $0.062$ \\
\hline\hline
\end{tabular}
\label{tab:tol_acp_sys}
\end{center}
\end{table}

%% file: contents/Summary_Conclusion.tex
\section{Summary and Conclusion}
\label{sec:summary}
We report measurements of branching fractions and {\it CP}-violating asymmetries of $B^+ \to h^+ \pi^0$ decays using data collected by the Belle II experiment in 2019 and 2020 at the $\Upsilon (4S)$ resonance, corresponding to 62.8 $\text{fb}^{-1}$ of integrated luminosity. The branching fractions and direct {\it CP}-violating asymmetries are measured to be 
\begin{center}
$\mathcal{B}(B^+ \to K^+\pi^0) = [11.9 ^{+1.1}_{-1.0}(\rm stat) \pm 1.6(\rm syst)]\times 10^{-6}$,
\end{center}
\begin{center}
$\mathcal{B}(B^+ \to \pi^+\pi^0) = [5.5 ^{+1.0}_{-0.9}(\rm stat) \pm 0.7(\rm syst)]\times 10^{-6}$,
\end{center}
\begin{center}
$\mathcal{A}_{\it CP}(B^+ \to K^+\pi^0) = -0.09 \pm 0.09 (\rm stat)\pm 0.03(\rm syst)$, and
\end{center}
\begin{center}
$\mathcal{A}_{\it CP}(B^+ \to \pi^+\pi^0) = -0.04 \pm 0.17 (\rm stat)\pm 0.06(\rm syst)$.
\end{center}

The results described above supersede the previous Belle~II results reported in Ref~\cite{Charmless:2020} using more data and an improved analysis that implements two-dimensional $M_{bc}$-$\Delta E$ fits and a more refined treatment of systematic uncertainties, and are in agreement with world averages~\cite{PDG}.

%% file: contents/acknowledgements.tex
\section*{Acknowledgments}
We thank the SuperKEKB group for the excellent operation of the
accelerator; the KEK cryogenics group for the efficient
operation of the solenoid; and the KEK computer group for
on-site computing support.
This work was supported by the following funding sources:
Science Committee of the Republic of Armenia Grant No. 18T-1C180;
Australian Research Council and research grant Nos.
DP180102629, 
DP170102389, 
DP170102204, 
DP150103061, 
FT130100303, 
and
FT130100018; 
Austrian Federal Ministry of Education, Science and Research,
Austrian Science Fund No. P 31361-N36, and
Horizon 2020 ERC Starting Grant no. 947006 ``InterLeptons''; 
Natural Sciences and Engineering Research Council of Canada, Compute Canada and CANARIE;
Chinese Academy of Sciences and research grant No. QYZDJ-SSW-SLH011,
National Natural Science Foundation of China and research grant Nos.
11521505,
11575017,
11675166,
11761141009,
11705209,
and
11975076,
LiaoNing Revitalization Talents Program under contract No. XLYC1807135,
Shanghai Municipal Science and Technology Committee under contract No. 19ZR1403000,
Shanghai Pujiang Program under Grant No. 18PJ1401000,
and the CAS Center for Excellence in Particle Physics (CCEPP);
the Ministry of Education, Youth and Sports of the Czech Republic under Contract No.~LTT17020 and 
Charles University grants SVV 260448 and GAUK 404316;
European Research Council, 7th Framework PIEF-GA-2013-622527, 
Horizon 2020 Marie Sklodowska-Curie grant agreement No. 700525 `NIOBE,' 
and
Horizon 2020 Marie Sklodowska-Curie RISE project JENNIFER2 grant agreement No. 822070 (European grants);
L'Institut National de Physique Nucl\'{e}aire et de Physique des Particules (IN2P3) du CNRS (France);
BMBF, DFG, HGF, MPG, and AvH Foundation (Germany);
Department of Atomic Energy under Project Identification No. RTI 4002 and Department of Science and Technology (India);
Israel Science Foundation grant No. 2476/17,
United States-Israel Binational Science Foundation grant No. 2016113, and
Israel Ministry of Science grant No. 3-16543;
Istituto Nazionale di Fisica Nucleare and the research grants BELLE2;
Japan Society for the Promotion of Science,  Grant-in-Aid for Scientific Research grant Nos.
16H03968, 
16H03993, 
16H06492,
16K05323, 
17H01133, 
17H05405, 
18K03621, 
18H03710, 
18H05226,
19H00682, 
26220706,
and
26400255,
the National Institute of Informatics, and Science Information NETwork 5 (SINET5), 
and
the Ministry of Education, Culture, Sports, Science, and Technology (MEXT) of Japan;  
National Research Foundation (NRF) of Korea Grant Nos.
2016R1\-D1A1B\-01010135,
2016R1\-D1A1B\-02012900,
2018R1\-A2B\-3003643,
2018R1\-A6A1A\-06024970,
2018R1\-D1A1B\-07047294,
2019K1\-A3A7A\-09033840,
and
2019R1\-I1A3A\-01058933,
Radiation Science Research Institute,
Foreign Large-size Research Facility Application Supporting project,
the Global Science Experimental Data Hub Center of the Korea Institute of Science and Technology Information
and
KREONET/GLORIAD;
Universiti Malaya RU grant, Akademi Sains Malaysia and Ministry of Education Malaysia;
Frontiers of Science Program contracts
FOINS-296,
CB-221329,
CB-236394,
CB-254409,
and
CB-180023, and SEP-CINVESTAV research grant 237 (Mexico);
the Polish Ministry of Science and Higher Education and the National Science Center;
the Ministry of Science and Higher Education of the Russian Federation,
Agreement 14.W03.31.0026;
University of Tabuk research grants
S-0256-1438 and S-0280-1439 (Saudi Arabia);
Slovenian Research Agency and research grant Nos.
J1-9124
and
P1-0135; 
Agencia Estatal de Investigacion, Spain grant Nos.
FPA2014-55613-P
and
FPA2017-84445-P,
and
CIDEGENT/2018/020 of Generalitat Valenciana;
Ministry of Science and Technology and research grant Nos.
MOST106-2112-M-002-005-MY3
and
MOST107-2119-M-002-035-MY3, 
and the Ministry of Education (Taiwan);
Thailand Center of Excellence in Physics;
TUBITAK ULAKBIM (Turkey);
Ministry of Education and Science of Ukraine;
the US National Science Foundation and research grant Nos.
PHY-1807007 
and
PHY-1913789, 
and the US Department of Energy and research grant Nos.
DE-AC06-76RLO1830, 
DE-SC0007983, 
DE-SC0009824, 
DE-SC0009973, 
DE-SC0010073, 
DE-SC0010118, 
DE-SC0010504, 
DE-SC0011784, 
DE-SC0012704; 
and
the National Foundation for Science and Technology Development (NAFOSTED) 
of Vietnam under Grant No. 103.99-2020.50.

%% file: contents/Non_Signal_Enhance.tex
\section{Non-signal enhanced plots of the fits}
\label{app:Non}
\begin{figure}[htb]
\centering
\includegraphics[width=0.475\textwidth]{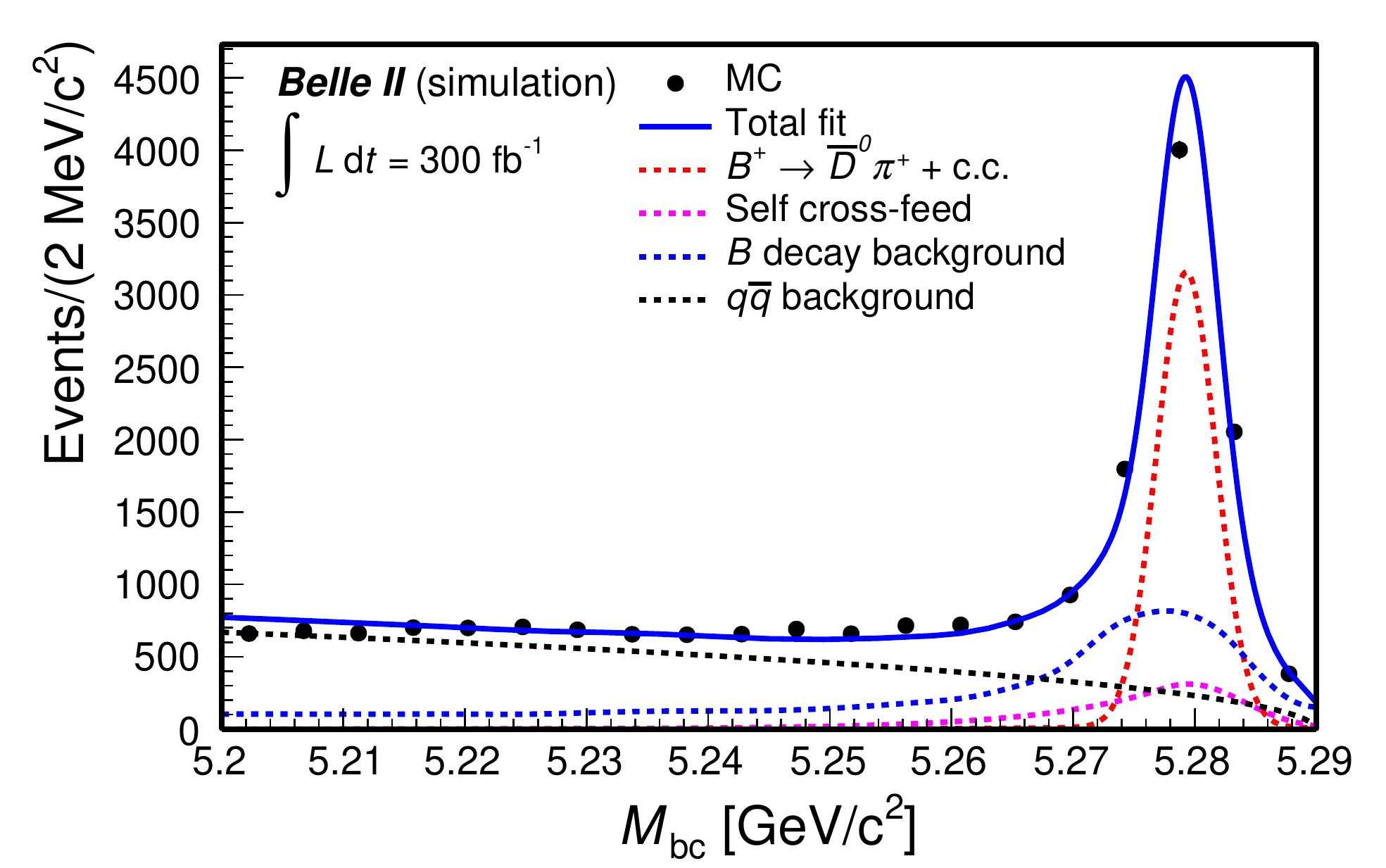}
\includegraphics[width=0.475\textwidth]{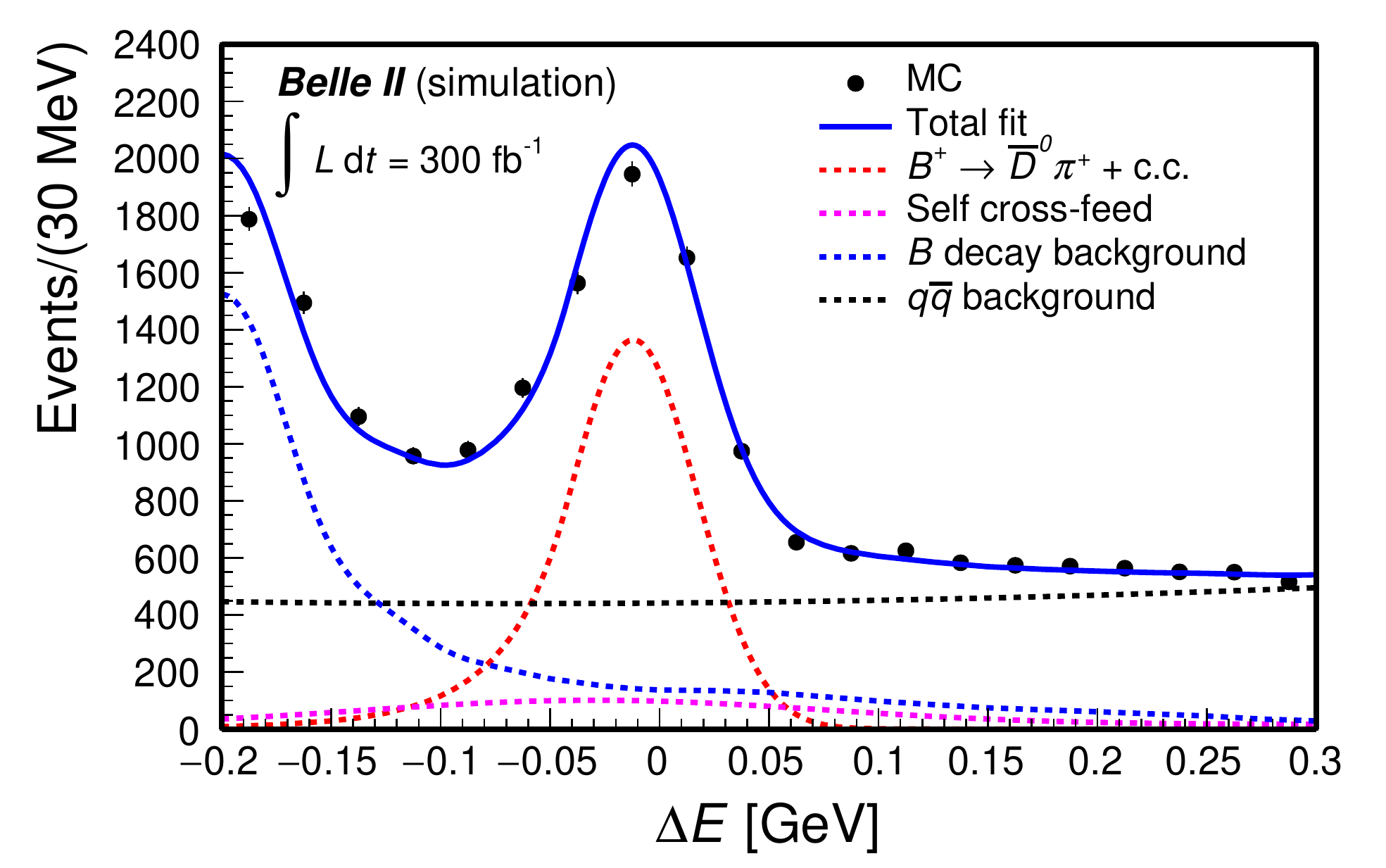}
\includegraphics[width=0.475\textwidth]{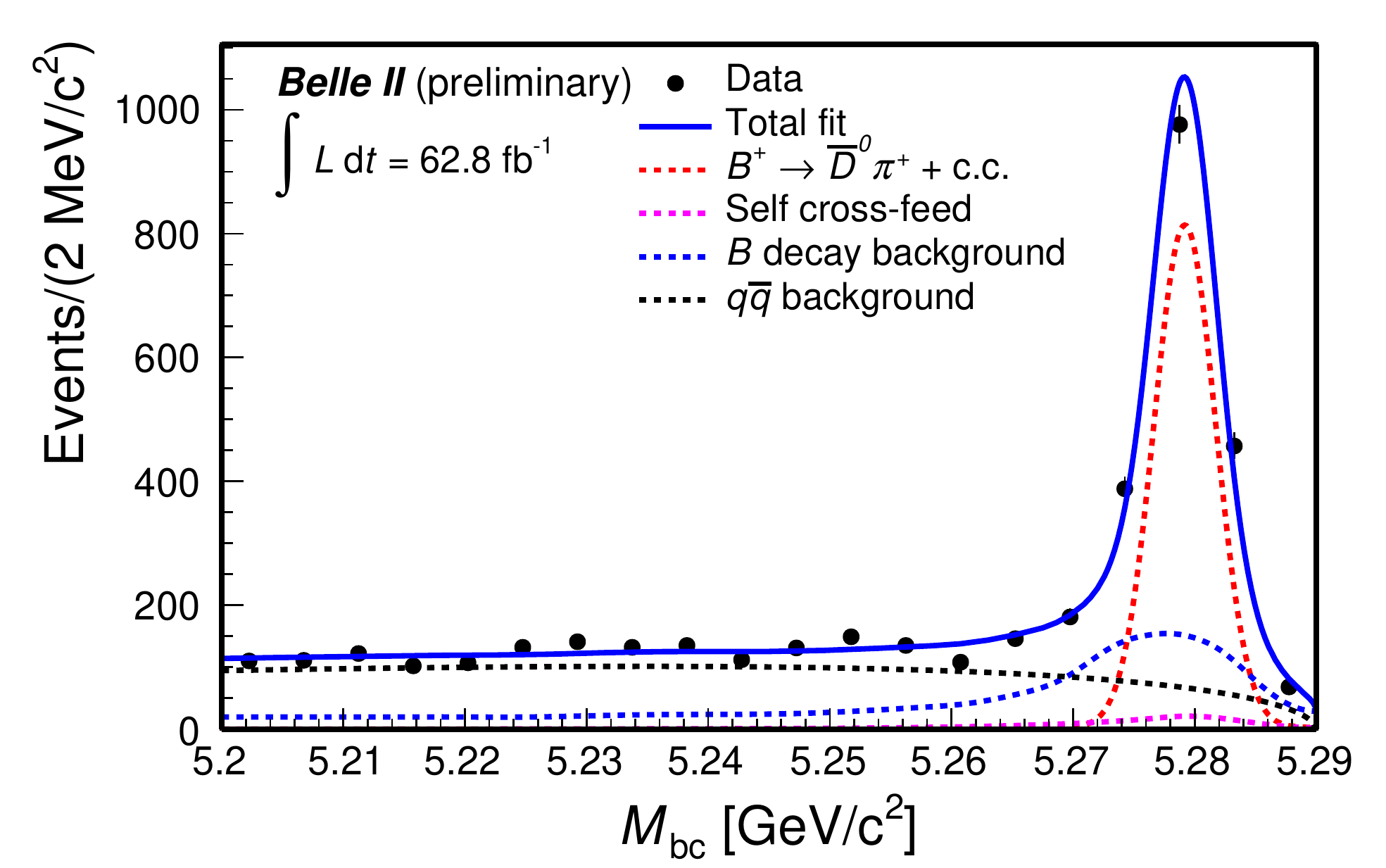}
\includegraphics[width=0.475\textwidth]{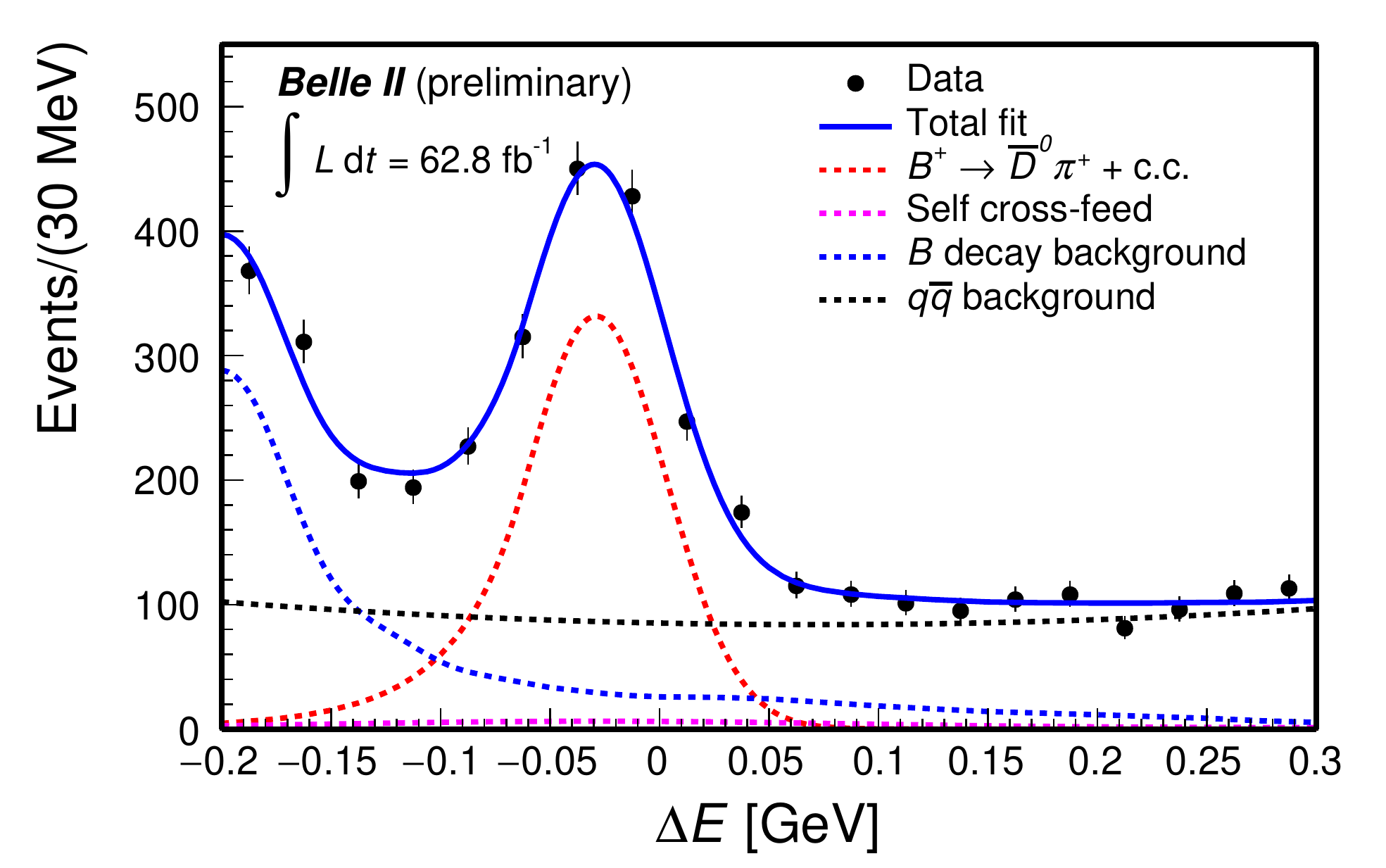}
\caption{Distributions of (left) $M_{bc}$ and (right) $\Delta E$ for $B^+\to \overline{D}^0 (\to K^+\pi^-\pi^0)\pi^+$ candidates reconstructed in (top) simulated events and (bottom) real data selected with an optimized continuum-suppression requirement. The projections of the fit are overlaid.}
\label{fig:CR_fitting_results_extra}
\end{figure}
\begin{figure}[htb]
\centering
\includegraphics[width=0.475\textwidth]{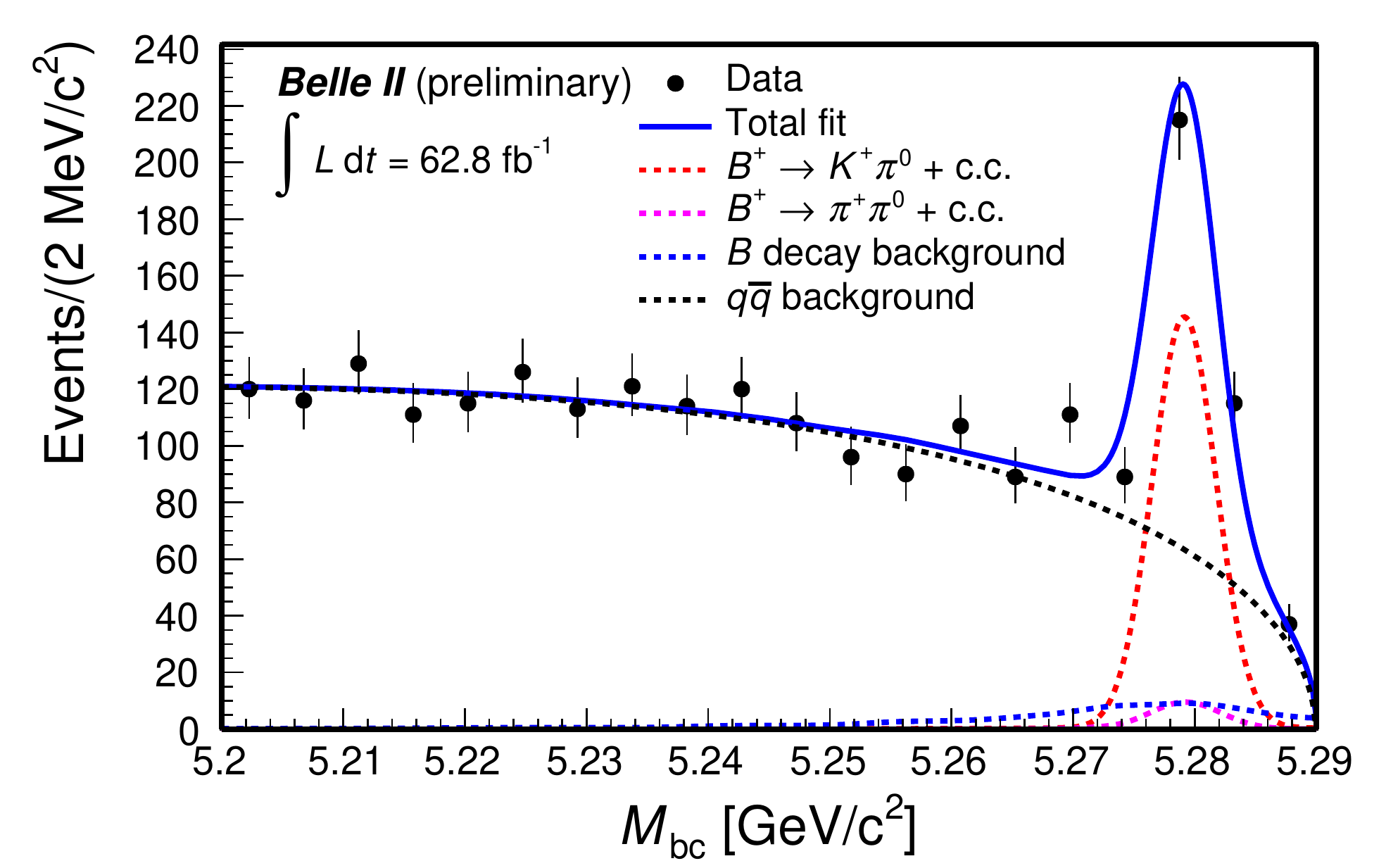}
\includegraphics[width=0.475\textwidth]{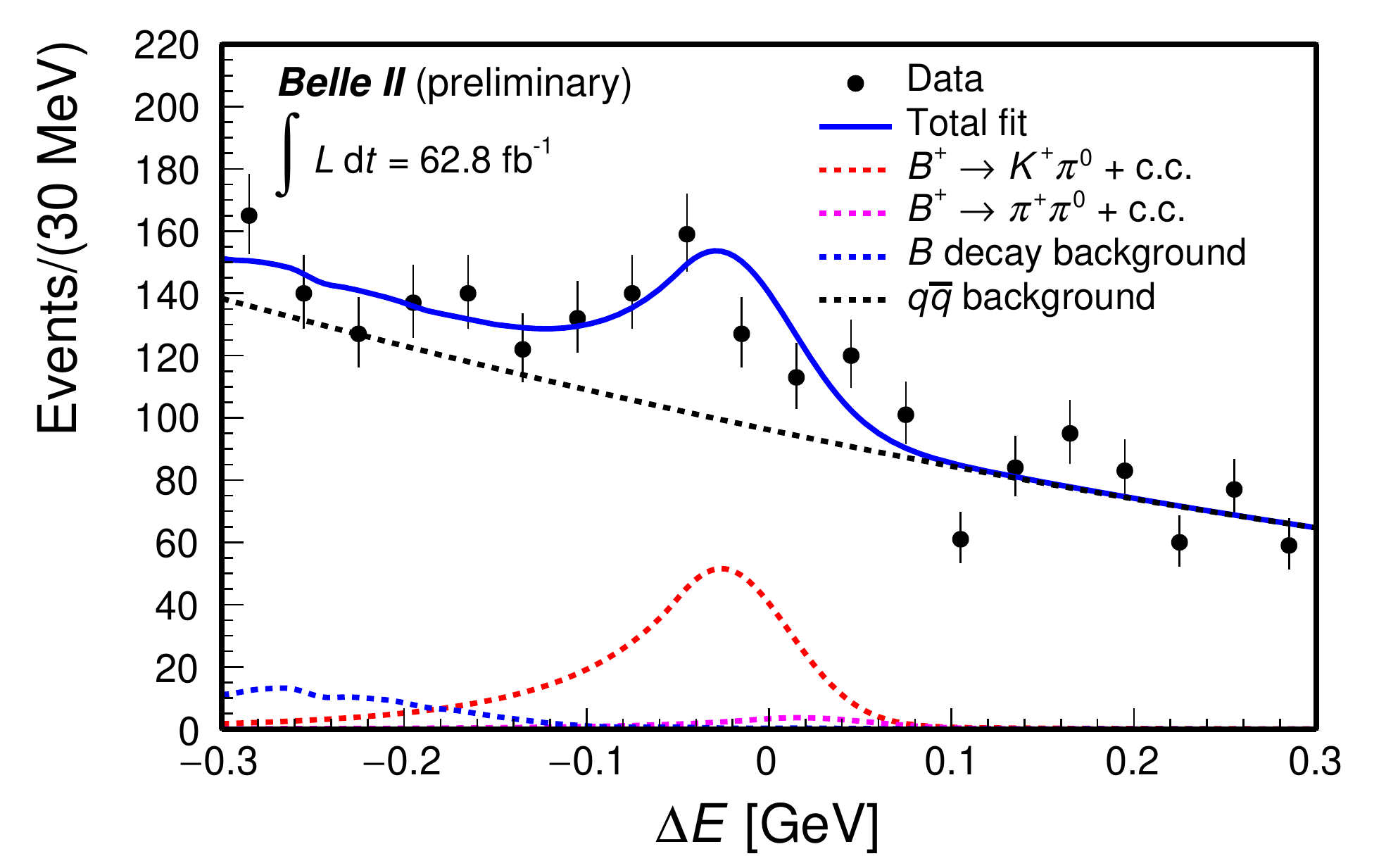}\\
{a) $B^+\to K^+\pi^0$}\vspace{2mm}\\
\includegraphics[width=0.475\textwidth]{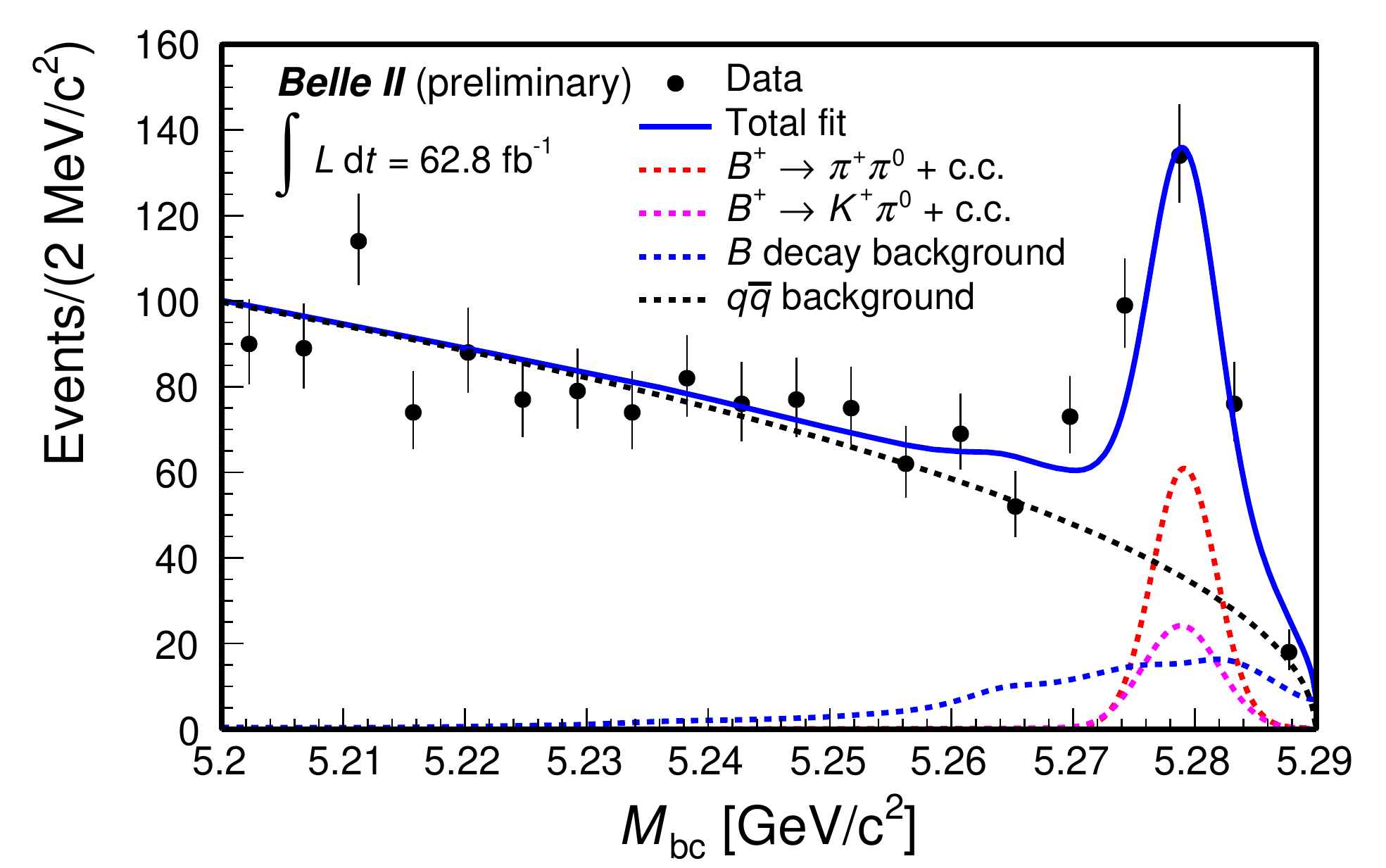}
\includegraphics[width=0.475\textwidth]{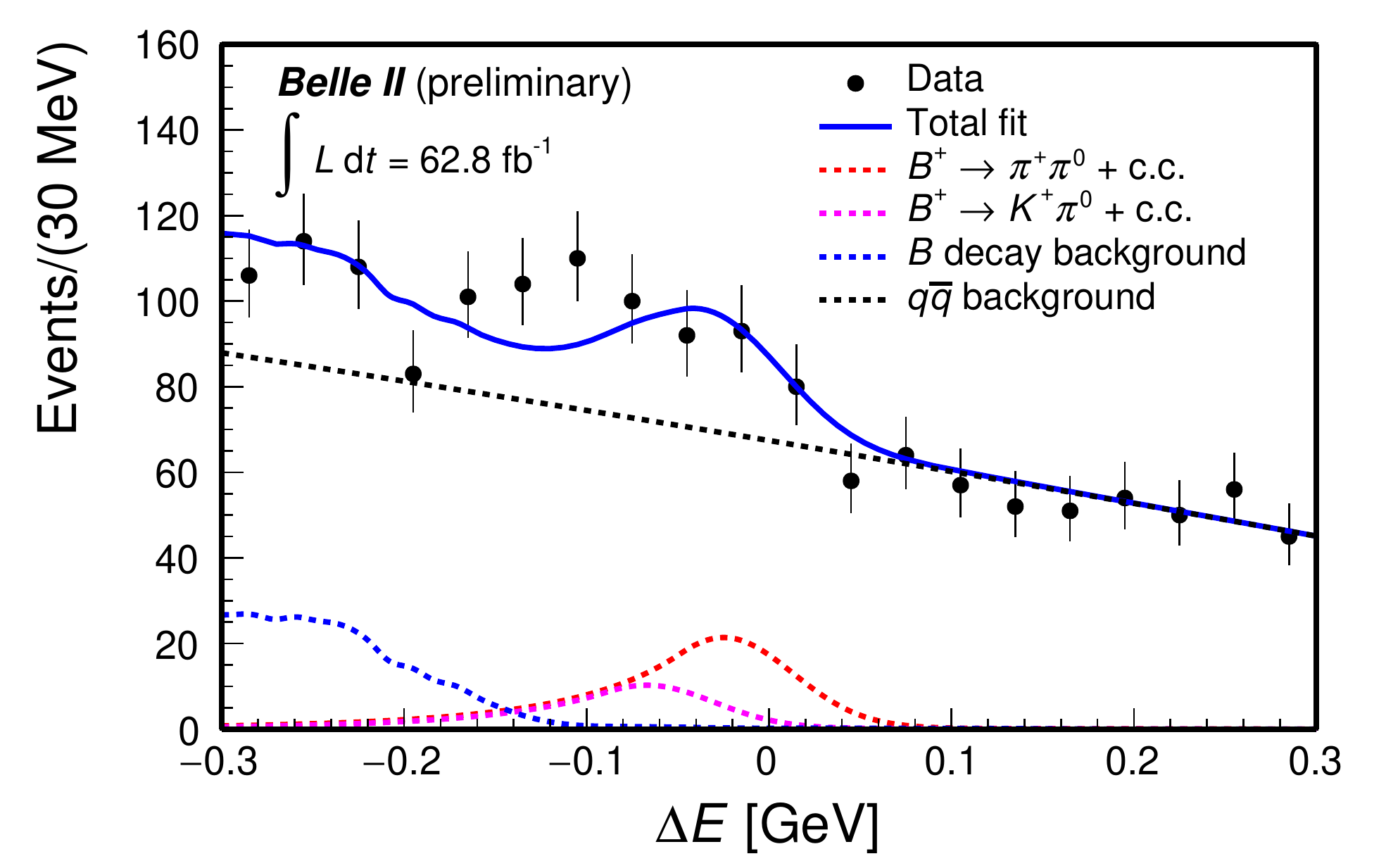}\\
{b) $B^+\to \pi^+\pi^0$}\vspace{2mm}\\
\caption{Distributions of (left) $M_{bc}$ and (right) $\Delta E$ for $B^+ \to h^+ \pi^0$ candidates reconstructed in 2019–2020 Belle~II data selected with an optimized continuum-suppression requirement. The projections of the fit are overlaid.}
\label{fig:fitting_rs_data_moriond_extra}
\end{figure}
\begin{figure}[htb]
\centering
\includegraphics[width=0.475\textwidth]{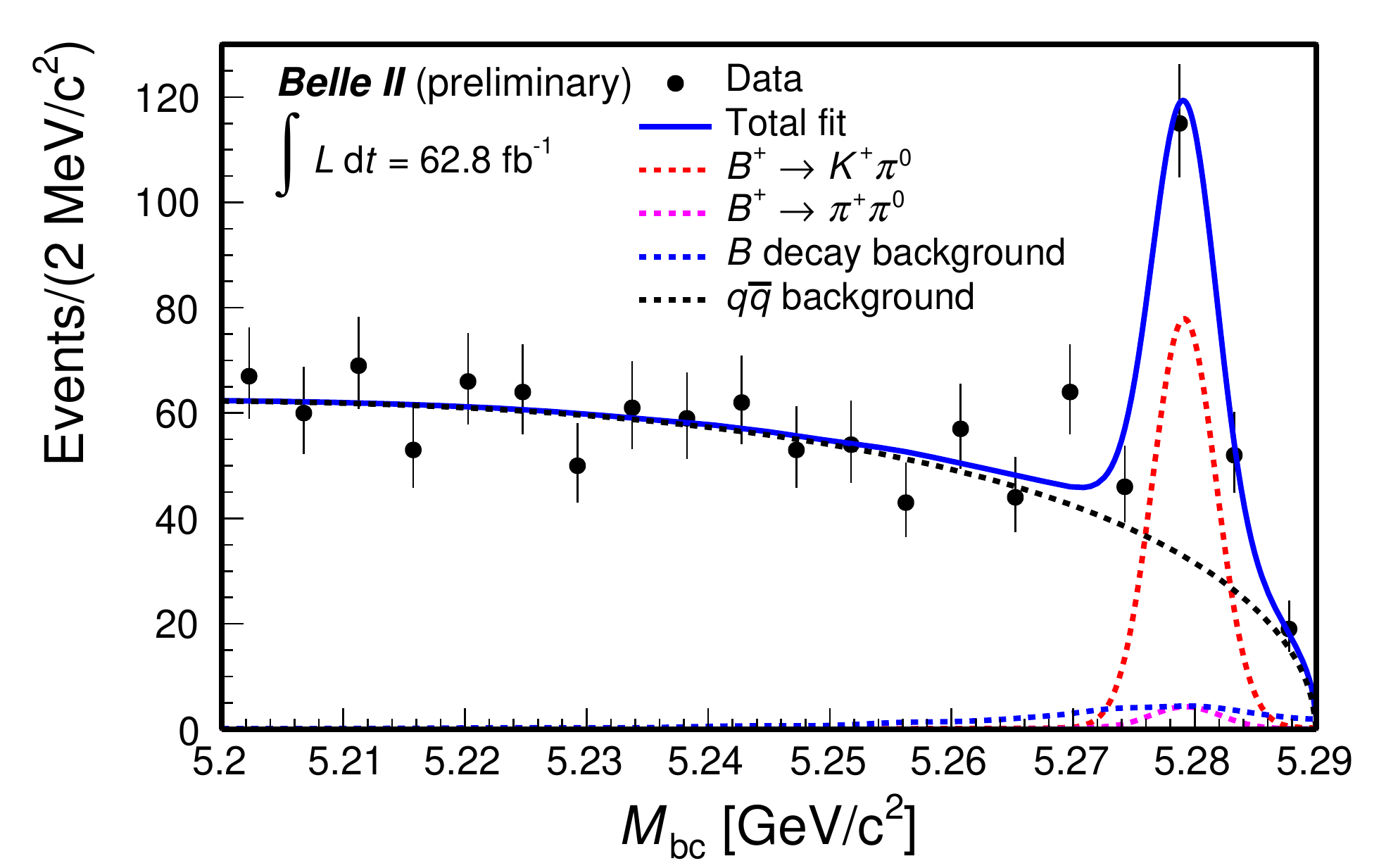}
\includegraphics[width=0.475\textwidth]{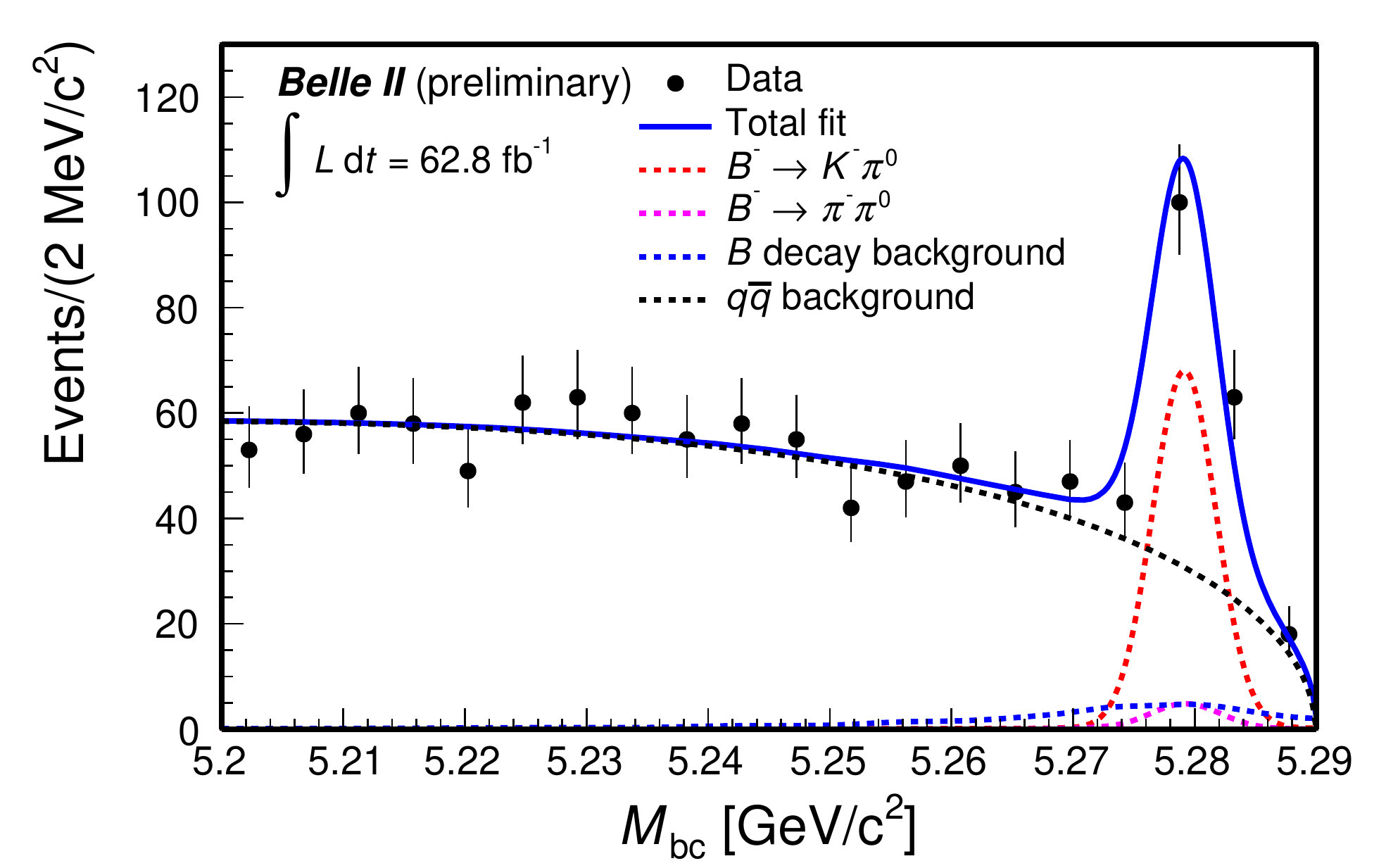}
\includegraphics[width=0.475\textwidth]{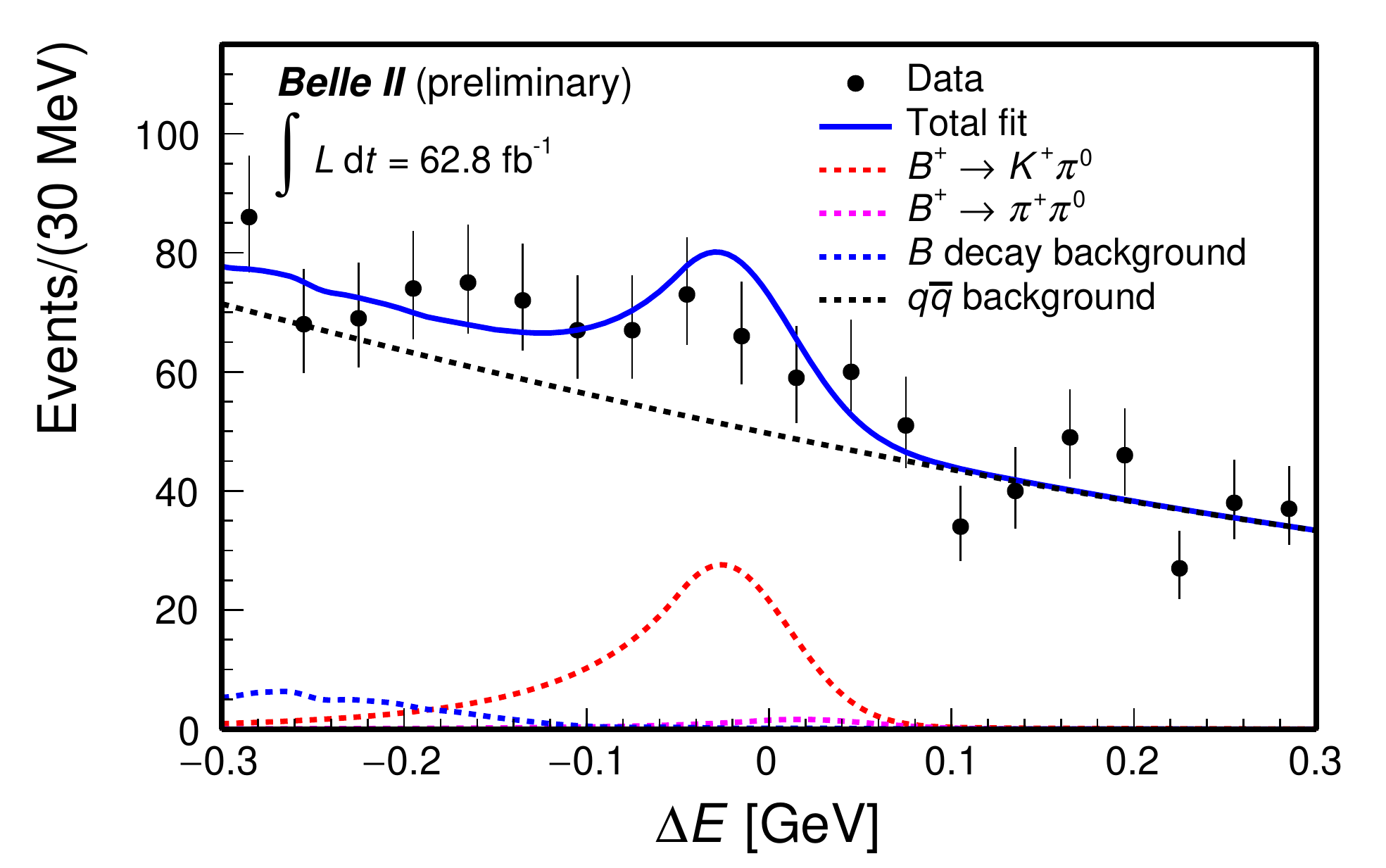}
\includegraphics[width=0.475\textwidth]{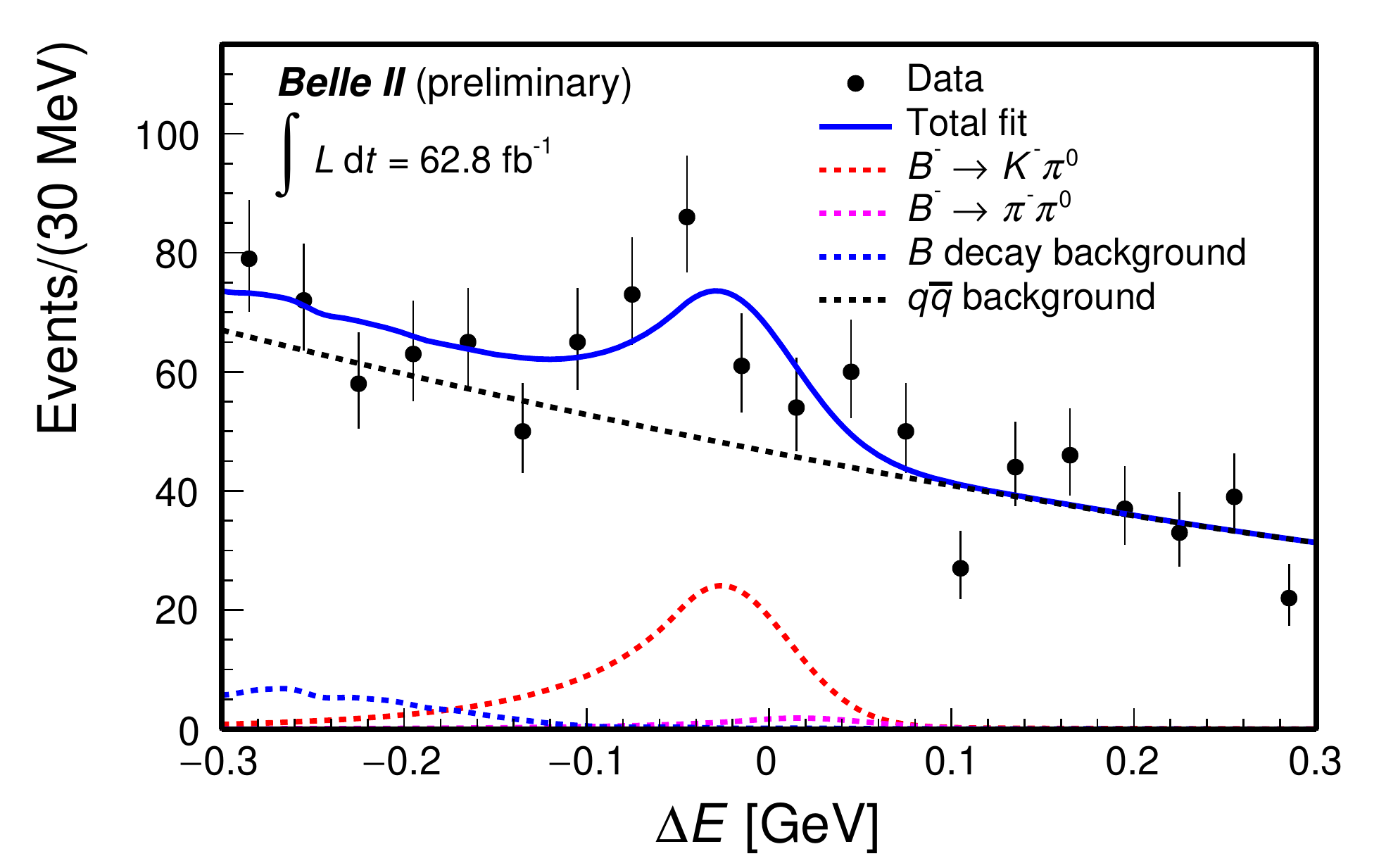}
\caption{Charge-specific distributions of (top) $M_{bc}$ and (bottom) $\Delta E$ for (left) $B^+ \to K^+ \pi^0$ and (right) $B^- \to K^- \pi^0$ candidates reconstructed in 2019–2020 Belle~II data selected with an optimized continuum-suppression requirement. The projections of the fit are overlaid.}
\label{fig:fitting_rs_Kpi_data_acp_moriond_extra}
\end{figure}
\begin{figure}[htb]
\centering
\includegraphics[width=0.475\textwidth]{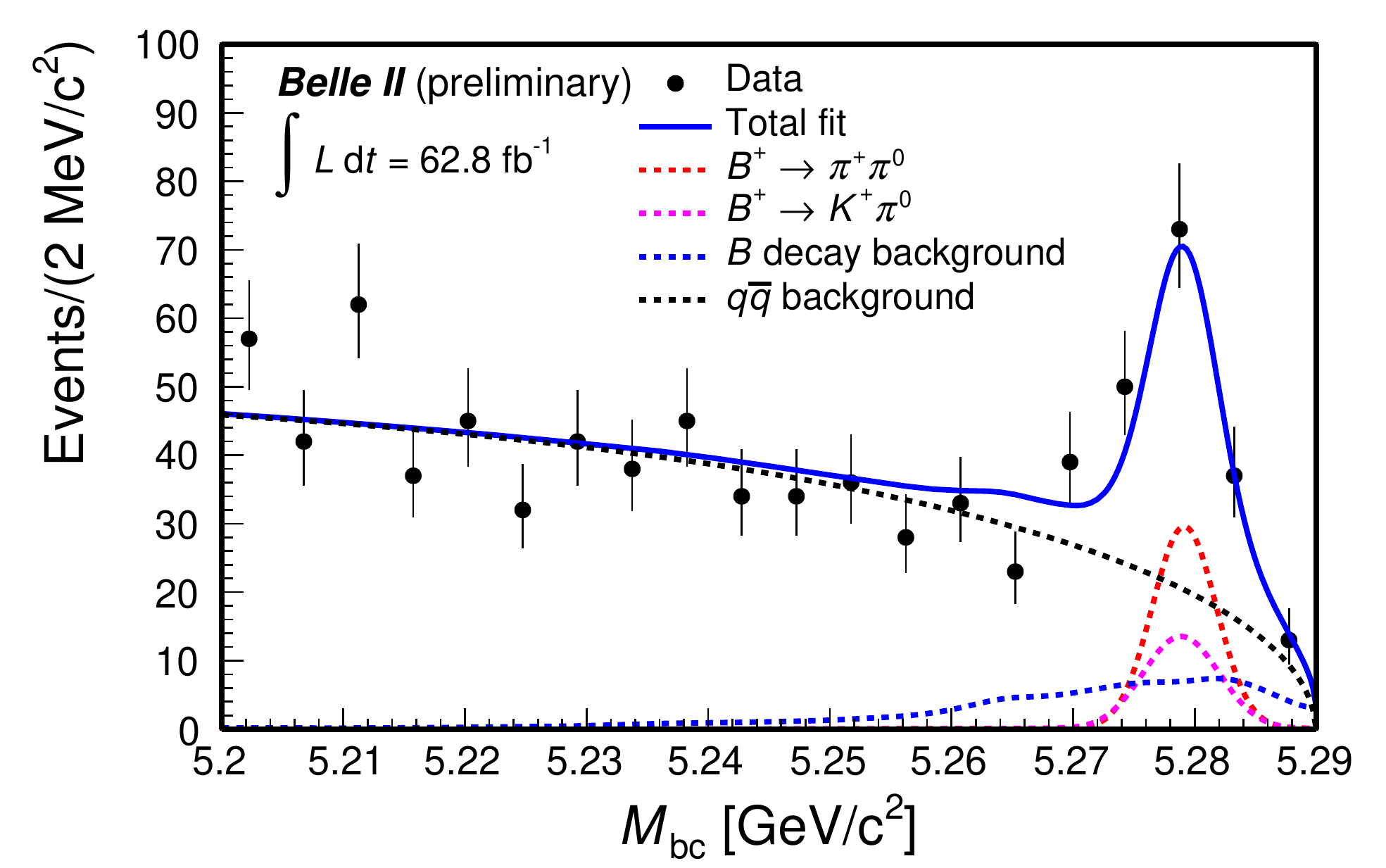}
\includegraphics[width=0.475\textwidth]{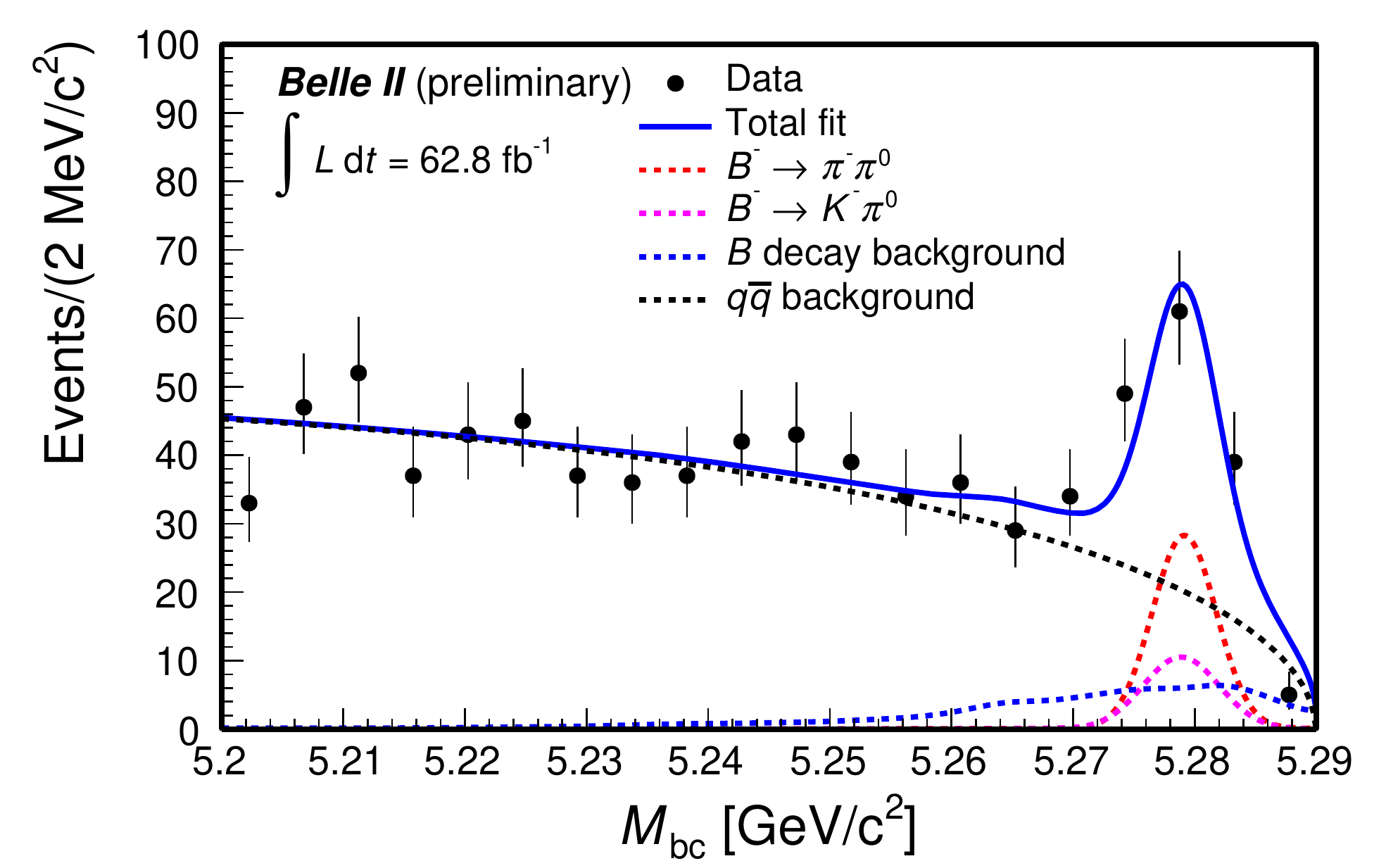}
\includegraphics[width=0.475\textwidth]{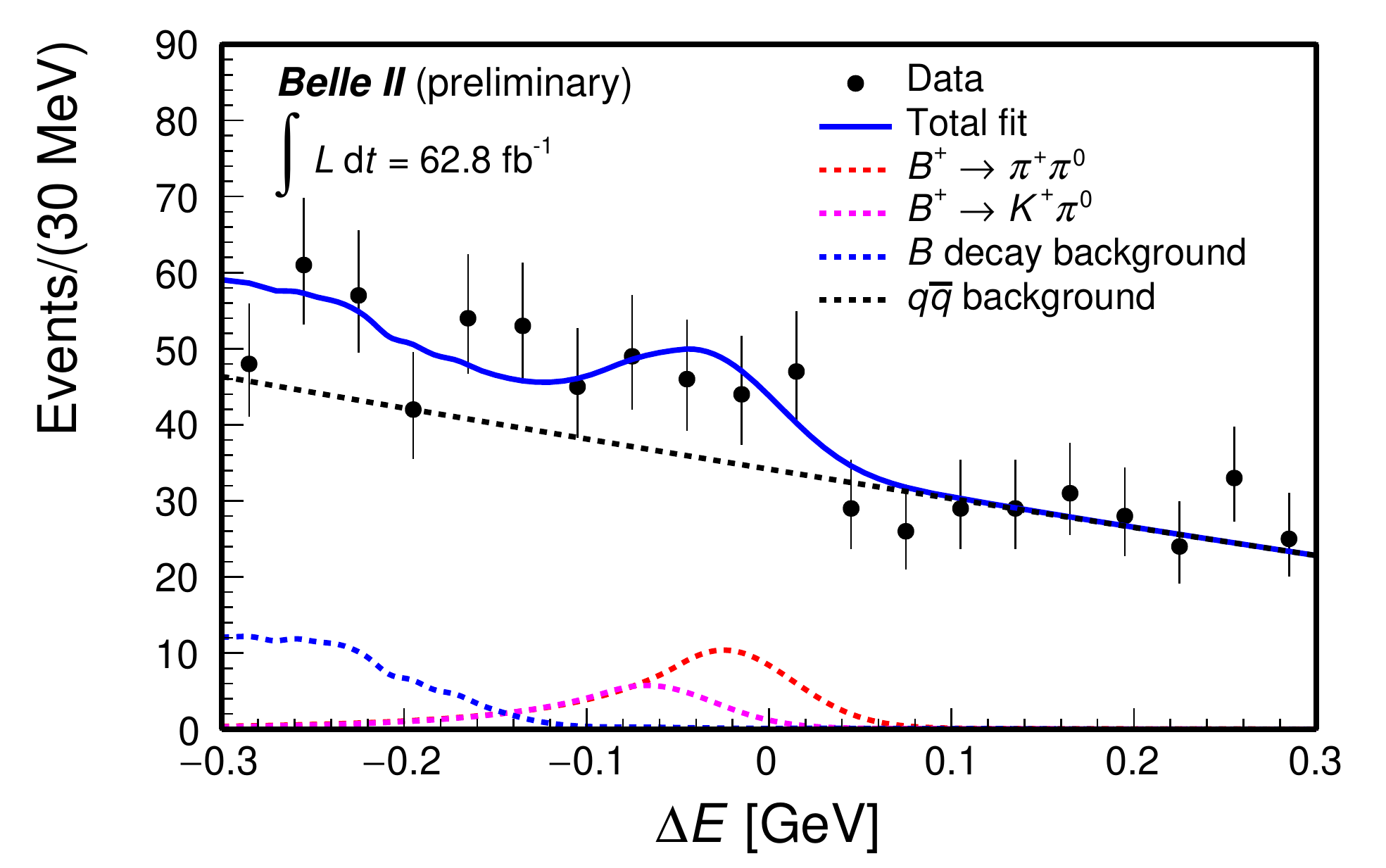}
\includegraphics[width=0.475\textwidth]{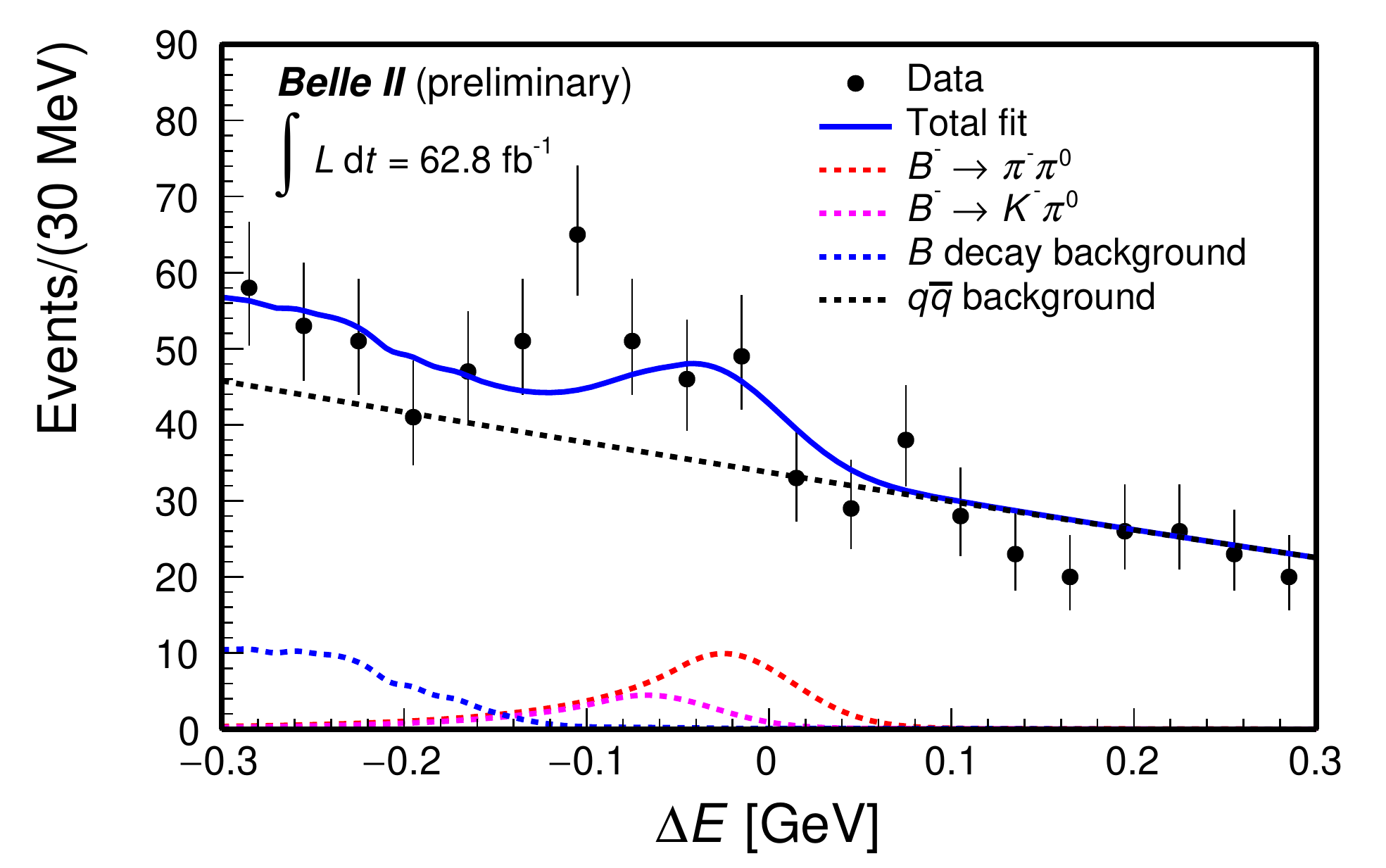}
\caption{Charge-specific distributions of (top) $M_{bc}$ and (bottom) $\Delta E$ for (left) $B^+ \to \pi^+ \pi^0$ and (right) $B^- \to \pi^- \pi^0$ candidates reconstructed in 2019–2020 Belle~II data selected with an optimized continuum-suppression requirement. The projections of the fit are overlaid.}
\label{fig:fitting_rs_pipi_data_acp_moriond_extra}
\end{figure}